\documentclass[sigconf]{acmart}
\renewcommand\footnotetextcopyrightpermission[1]{}
\AtBeginDocument{%
  \providecommand\BibTeX{{%
    \normalfont B\kern-0.5em{\scshape i\kern-0.25em b}\kern-0.8em\TeX}}}

\usepackage{array}
\usepackage{tabularx}
\def\aditya#1{\textcolor{red}{Aditya: #1}}
\def\aditya#1{}
\def\skipnoindent{\vskip0.1in\noindent}
\def\mlterm#1{\bf #1}
\def\ignoretext#1{}

\begin{document}

\title[Machine Learning for Intrusion Detection in Industrial Control Systems]{Machine Learning for Intrusion Detection in Industrial Control Systems: Applications, Challenges, and Recommendations}




\author{Muhammad Azmi Umer}
\affiliation{%
  \institution{DHA Suffa University}
  \institution{Karachi Institute of Economics and Technology}
}
\email{muhammadazmiumer@yahoo.com}

\author{Khurum Nazir Junejo}
\affiliation{%
  \institution{DNNae Inc.}
}
\email{junejo@gmail.com}

\author{Muhammad Taha Jilani}
\affiliation{%
  \institution{Karachi Institute of Economics and Technology}
}
\email{m.taha@kiet.edu.pk}

\author{Aditya P. Mathur}
\affiliation{%
  \institution{Singapore University of Technology and Design}
}
\email{aditya\_mathur@sutd.edu.sg}

\begin{abstract}

Methods from machine learning   are being applied to design   Industrial Control Systems resilient to cyber-attacks. Such methods  focus on two major areas: the detection of intrusions at the network-level  using the information acquired through network packets, and detection  of anomalies at the physical process level using data that represents the physical behavior of the system. This survey focuses  on four types of methods from machine learning in use for intrusion and anomaly detection,  namely,  supervised, semi-supervised, unsupervised, and reinforcement learning. Literature  available in the public domain was carefully selected, analyzed, and placed in a 7-dimensional space for ease of comparison. The survey is targeted at researchers, students, and practitioners.  Challenges associated in using the methods and research gaps are identified and recommendations are made to fill the gaps. 
\end{abstract}





\maketitle

\section{Introduction}

This article is a survey of methods from machine learning (ML) that are being applied to detect intrusions, or anomalies, in systems. The systems of interest in this survey are primarily those where an Industrial Control System (ICS) is used to control a physical process. Such systems are constituents of critical infrastructure in a city and country, and include the electric power grid, water treatment and distributions systems, and oil refineries. Such systems are a subset of a broader class of systems known as Cyber-Physical Systems (CPS) that  consist of   cyber  and physical subsystems.  These subsystems are integrated via sensors, actuators, and communications links to enable the control of the underlying physical process\,\cite{barbosa2016cross, rajkumar2012cyber, baheti2011cyber}. While ICS remain the focus of this survey, we have not avoided references to systems that do not use ICS, but fall in the CPS category.

\skipnoindent {\em Industrial Control Systems}: ICS  include a Supervisory Control and Data Acquisition (SCADA) system, Programmable Logic Controllers (PLCs), Remote I/O (RIO) units, sensors, and actuators. While the specific brand and types of such  subsystems may differ, their overall function is to effectively control the underlying physical process.  Successful and   unsuccessful attempts to affect the behavior of ICS has led to an  increase in research aimed at developing methods and tools to protect plants from malicious actors\,\cite{icsincident,icsCERTAdvisory}.  Such attempts by malicious actors are made possible, and are sometimes successful,  due to a variety of reasons including inadequate physical and or cyber protective measures and  network connectivity.  

\skipnoindent{\em Attacks on ICS}: Data in Table~\ref{tab:icsincident} is indicative of the rise in successful cyber-attacks on ICS.  A uranium enrichment plant in  Iran was attacked\,\cite{falliere2011w32} resulting in an increase in the failure of centrifuges. The  Maroochy water services were attacked by an ex-employee and a large quantity of sewage spilled into  a local park\,\cite{slay2008lessons}. A water treatment plant in the U.S. was attacked in 2006\,\cite{cardenas2011attacks}. Such attacks, and their impact, has led to a realization that new methods and tools, beyond the traditional mechanism, e.g., firewalls that protect communication networks,  are needed to protect ICS.   

\skipnoindent {\em Target audience}: Given an increasing body of literature focusing on using ML for defending ICS against cyber-attacks, it is important to subject this body of work from a critical perspective for the benefit of  researchers, students and practitioners.  Researchers and students aiming to explore the use of ML in defending ICS against cyber-attacks stand to benefit from this survey as it would allow them to identify gaps in the literature and weaknesses of existing methods. Practitioners, aiming to develop commercial tools for use in operational plants, stand to benefit from this survey as it would help them identify the most promising methods on which to base their tools. 

\skipnoindent{\em Keeping the survey live}: Given the rate at which research is progressing in the application of machine learning to detect cyber intrusions, it is likely that this survey will rapidly be rendered incomplete, or even outdated, soon after its publication. To ensure that the survey remains up-to-date, we have created a web site\footnote[1]{https://sites.google.com/view/crcsweb/survey-paper} where we will add new literature in this area with suitable comments. Tables in this article that place each research publication in a 7-dimensional space will be kept at this site and updated regularly.

\skipnoindent{\em Abbreviations and nomenclature}: Given the focus of this survey, the terms ``plant," ``system,"  and ICS are used synonymously. Such usage is justifiable as an ICS is a subsystem in a physical system and, when attacked, it impacts the underlying process, e.g., water filtration or uranium enrichment. We note that ICS enabled systems are Cyber-Physical Systems. However, as much as possible, we have avoided the use of the term CPS due its breadth and the fact that literature surveyed here focuses mostly on plants  controlled by an ICS. Literature related to detection of anomalies in network traffic is generally classified under   ``Intrusion detection" category. However, literature in the ICS domain that focuses on physical processes in a plant, is classified under ``anomaly detection." In this survey we use the  ``intrusion detection" to refer to anomaly detection in physical plants as well as the detection of network intrusions. Techniques from machine learning are often referred to by their abbreviations, e.g., RNN for Recurrent Neural Networks. This survey uses a large number of such abbreviations. To make it easy for a reader new to  machine learning each abbreviation used in this article, and its expansion,  is listed in alphabetic order in Table~\ref{tab:abbreviations} placed at the end of this survey. 
\begin{table*}[h]
\caption{Incidents on Industrial Control Systems}
\label{tab:icsincident}
\begin{tabular}{|l|l|l|l|}
\hline
\multicolumn{1}{|c|}{\textbf{Year}} & \multicolumn{1}{c|}{\textbf{Incident}} & \multicolumn{1}{c|}{\textbf{Year}} & \multicolumn{1}{c|}{\textbf{Incident}} \\ \hline
2019 & LockerGoga Ransomware\,\cite{lockergoga} & 2014 & Port Hudson Paper Mill Insider Threat\,\cite{hudson} \\ \hline
2018 & Olympic Destroyer\,\cite{olympic} & 2013 & Havex\,\cite{havex} \\ \hline
2018 & TRITON Triconex SIS Malfunction\,\cite{triton} & 2012 & Shamoon\,\cite{Shamoon} \\ \hline
2017 & TEMP.isotope Campaign\,\cite{isotope} & 2011 & Duqu\,\cite{duqu} \\ \hline
2017 & BadRabbit Ransomware\,\cite{rabbit} & 2010 & Stuxnet\,\cite{falliere2011w32} \\ \hline
2017 & EternalPetya Ransomware\,\cite{petya} & 2008 & CIA Reports Foreign Utilities Hacked\,\cite{cia} \\ \hline
2017 & WannaCry Ransomware\,\cite{cry} & 2007 & Aurora Generator Test\,\cite{aurora} \\ \hline
2016 & Industroyer Ukraine Blackout\,\cite{industryukraine} & 2003 & Northeast Blackout\,\cite{northeast} \\ \hline
2015 & BlackEnergy 3 Ukraine Blackout\,\cite{ukraineBlackout} & 2001 & Maroochy Sewage Spill\,\cite{slay2008lessons} \\ \hline
\end{tabular}
\end{table*}
\skipnoindent{\em Organization}:  The remainder of this survey article is organized as follows.  In Section~\ref{sec:intrusionDetectionSystems} we introduce Intrusion Detection Systems (IDS) and categorize them broadly. A large number of articles had to be collected for this survey to be possible. The collection process is summarized in Section~\ref{sec:collection}. There are other surveys reported that also focus on ML techniques as applied to ICS. Such surveys are cited with differences from our survey identified in Section~\ref{sec:relatedSurveys}. The literature surveyed and evaluated is placed in a 7-dimensional space described in Section~\ref{sec:dimensions}. Various methods from machine learning used for intrusion detection are categorized and explained in Section~\ref{sec:MLApproaches}. This is followed by Sections~\ref{sec:supervisedLearning}, \ref{sec:unsupervisedLearning}, \ref{sec:semisupervisedLearning}, and \ref{sec:reinforcementLearning} where we examine, respectively, the literature that focuses on the use of supervised, unsupervised, semi-supervised, and reinforcement learning for intrusion detection. Major challenges and recommendations related to IDS in ICS are discussed in Section~\ref{sec:challenges}. Section \ref{sec:conclusions} has summarized the overall work and discussed the conclusion.


\section{Intrusion Detection Systems}
\label{sec:ids} \label{sec:intrusionDetectionSystems}

Before diving into a detailed survey, we summarize below the various types of intrusion detection systems (IDS). Such systems aim at detecting intrusions and anomalies  during plant operation. The detected intrusions and anomalies are reported to plant engineers who are then expected to take appropriate actions to prevent undesirable consequences such as service disruption and component damage. Three types of IDS are considered in the following, namely, signature-based, specification-based, and behavior based.

\begin{figure*}[h]
\centering
\includegraphics[width=1.0\linewidth, height=8.5cm]{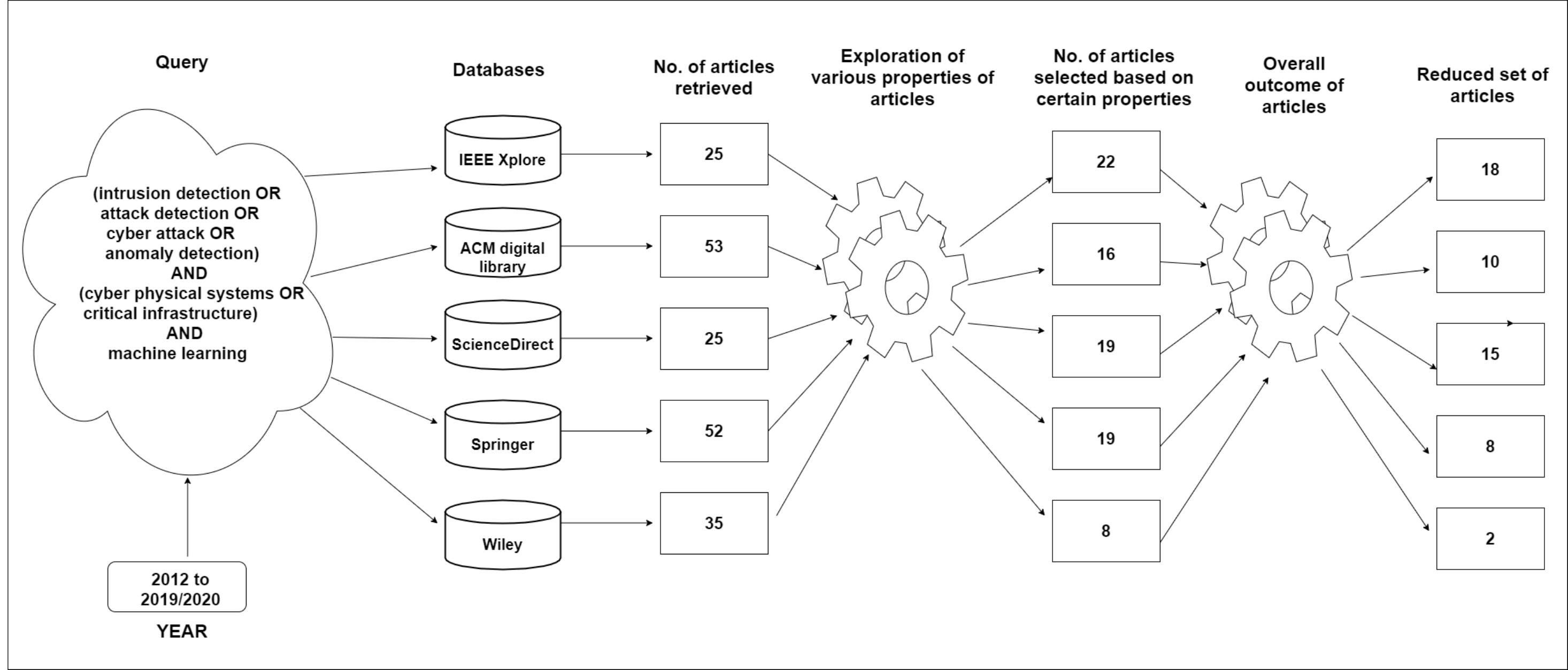}
\caption{Retrieval and Selection of Articles}
\label{fig:articles}
\end{figure*}

\subsection {Signature-based IDS}
This type of IDS requires a predefined dictionary of attack patterns. It detects an intrusion if any pattern detected during plant operation  matches one or more of the predefined attack patterns\,\cite{gao2014cyber}. Though this approach maintains a low rate of  false positives, it fails to detect zero-day attacks. Further, it is often difficult to produce an exhaustive dictionary of attack signature in complex physical processes. There are numerous ways to automate the generation of malware signatures. For example, in the study reported in\,\cite{nahmias2019trustsign}  malware signatures were generated in private cloud  using deep feature transfer learning.  Volatile memory dumps were extracted during the malware activity by querying the hypervisor of the virtual machine. Malicious processes were extracted from the memory dumps and  converted to images. Later, these images served as input to a pre-trained deep neural network model, namely, VGG19. The proposed model is robust and fast as it does not require training on new input data. However, as it generates signatures using only the available malware processes,  it could be prone to zero-day attacks.

\subsection{Specification-based IDS}
This approach develops a mathematical model to define the normal operation of the physical process under consideration. An anomaly is said to exist whenever the process deviates from the prediction by the predefined model\,\cite{mitchell2015behavior}. Such models are developed with the help of  experts and plant design. While the experts may have  knowledge of physical processes, there are issues related to the aging of the physical system,  inaccuracies that may exist in operational manuals, and interpretation of the process behavior. Secondly, it is  difficult to develop accurate mathematical models for complex distributed physical systems. The study reported in\,\cite{adepu2016using} derived the invariants (specifications) from the design of a water treatment plant. They used it to detect cyber-attacks on the plant. However, unless automated, the proposed approach is unable to derive the specifications of complex physical processes that are not reflected in the design document. A study reported in \cite{6133080} also used a specification-based approach for intrusion detection in Advanced Metering Infrastructures (AMI). They used sensors to monitor the traffic at meters and access points at the network, transport, and application layer. They made a set of specifications and policies to ensure the safety of meters and AMI, respectively. 

\subsection{Behavior-based IDS} 
This approach is based on the operational data from  the physical system. Based on  data collected, a model is trained on the normal and abnormal behavior of the process and used  to detect intrusions. This approach is favored against incorrect vendor specifications as it trains the model on empirical data\,\cite{ junejo2020predictive} and thus helps in identifying incorrect vendor specifications. For instance, a study reported in\,\cite{david2016CRC} noticed different levels of a water tank in a water treatment plant. According to the vendor specifications, the upper bound of on the volume of water in the  tank was 1100 liters; this value was also encoded in the control logic of the PLCs in the ICS. Analysis of data obtained through level sensors associated with the tank revealed that the upper bound in practice was 900 liters.

Traditional behavior-based approaches relied on statistical techniques\,\cite{ye2002multivariate} such as the mean and standard deviation of sensor   readings.  Lately, machine learning (ML)  techniques are being used extensively  as behavior-based approaches to secure  ICS. State of the art techniques using this approach have been reported in the literature. Such techniques are gaining popularity among researchers and commercial vendors mainly due to the availability of  high computing power and tools to detect the non-linear relations and unobserved regularities in the massive volumes of data. Nevertheless, there remain serious problems associated with these techniques including the detection of zero-day attacks, ensuring an acceptable rate of false alarms, and managing computational complexity. These problems are creating a bottleneck for the deployment of IDS based on these techniques, in particular in complex  Industrial Control Systems (ICS). This article discusses these techniques in detail within the paradigm of intrusion detection in ICS. It also discusses the associated problems and offers recommendations.

\section{Collection of Articles}
\label{coa}\label{sec:collection}
Apart from other relevant articles, a major set of articles reported in this study were collected using a systematic approach. Due to the inaccessibility of Web of Science and Scopus, five major databases including IEEE Xplore, ACM digital library, ScienceDirect, Springer, and Wiley were explored in-depth. Several queries were used to retrieve the relevant articles. These queries can be combined to form a single  query using logical connectives, as for example
\newline
\textit{(INTRUSION DETECTION \textbf{OR} ATTACK DETECTION \textbf{OR} CYBER ATTACK \textbf{OR} ANOMALY DETECTION) \textbf{AND} (CYBER PHYSICAL SYSTEMS \textbf{OR} CRITICAL INFRASTRUCTURE) \textbf{AND} MACHINE LEARNING.}

All  articles from 2012 to 2020 were retrieved in  multiple iterations as described in Figure\,\ref{fig:articles}. In the first iteration, a breadth-first study of each article was performed to extract various properties including the approach, limitations, strengths, etc. In the second iteration, articles were selected based on the relevance of the proposed approach to ICS. For example, some articles were related to IDS but did  not emphasize ICS or CPS, and hence were not selected for further analysis. Twenty-five articles were retrieved from IEEE Xplore. Here, the focus was only on journals and magazine articles. Fifty-three articles were retrieved from the ACM Digital Library. From this library, only the articles from journals and conferences of core rank A and B were selected. Twenty-five articles were retrieved from ScienceDirect. Nineteen articles were selected in the first iteration and fifteen in the second. Fifty-two articles were retrieved from Springer of which  nineteen articles were selected in the first iteration and eight  in the second. Thirty-five articles were retrieved from Wiley of which eight were selected in the first iteration and only two were selected in the second.


\section{Related Surveys}
\label{sec:relatedSurveys}

\begin{table*}
\caption{Comparison with Past Surveys}
\label{tab:pastsurveys}
\begin{tabular}{|l|p{4in}|}
\hline
{\bf Past Surveys}&{\bf Difference}\\
\hline
\cite{bhamare2019cybersecurity}&Main theme is to shift current ICS to cloud based infrastructure.\\
\cite{aleesa2019review}& Focus on IDS in general terms; not specifically on ICS.\\
\cite{luo2020deep}&Focus on Deep Learning (DL) techniques with types of anomalies, evaluation metrics, strategies, and implementation details; different  taxonomy\\
\cite{10.1145/3203245}&A general survey of physics-based attack detection in CPS; not focused on ML. \\
\cite{10.1145/2542049}& A survey of IDS in CPS focusing only on detection technique and audit material.\\
\cite{gupta2010networked,shi2011survey}& A survey of CPS  discussing challenges and future trends; does not focus on IDS approaches for CPS.\\
\cite{sommer2010outside,garcia2009anomaly,axelsson2000intrusion,anantvalee2007survey,chen2006survey}& Focus on Network-based IDS.\\
\cite{9152841}& Focus on Reinforcement Learning (RL) based Q-learning methods for securing CPS.\\
\cite{zhu2010scada}& Focus on SCADA specific intrusion detection and prevention.\\

\hline
\end{tabular}
\end{table*}

A comparison of related surveys is presented in Table \ref{tab:pastsurveys}. A survey of ICS security focusing mainly on ML is reported  in\,\cite{bhamare2019cybersecurity}. The article has discussed the benefits and shortcomings of using ML techniques for detecting anomalies in ICS. The need for shifting current ICS to cloud-based infrastructure was the main theme of this research. This survey has discussed minimal work on machine learning-based IDS. Also, only an overview of machine learning approaches was emphasized.

Deep Learning-based intrusion detection systems are discussed in\,\cite{aleesa2019review}. This work focuses on intrusion detection in its general terms, not focusing on ICS. The  work is divided into the frameworks, developed IDS, datasets, and testbeds. A survey of deep learning techniques for anomaly detection is reported in\,\cite{luo2020deep}. A taxonomy was  developed  for the survey which includes type of anomalies, evaluation metrics, strategies, and implementation details. 

A survey of physics-based anomaly detection is reported in\,\cite{10.1145/3203245}. The authors  developed a taxonomy to identify the key characteristics of their survey. This taxonomy consists of attack detection, attack location, and validation.  Attack detection is divided into prediction and detection statistics. Metrics and the implementation to verify and validate the performance of attack detection algorithms, are discussed. A survey of intrusion detection techniques is reported in\,\cite{10.1145/2542049}. This survey focuses on two dimensions, i.e., the audit material and  detection techniques. Apart from these two dimensions, the survey reported in the article here focuses on several other dimensions as well as discussed in section\,\ref{sec:IDS}.

A survey of CPS is reported in\,\cite{gupta2010networked,shi2011survey}. This survey discusses the challenges and future research trends but did not focus on IDS approaches for CPS. The network-based IDS was surveyed in\,\cite{sommer2010outside,garcia2009anomaly,axelsson2000intrusion,anantvalee2007survey,chen2006survey}, but the authors do not  address the  scenario which differs from conventional networks. A survey reported in\,\cite{9152841} focuses on reinforcement learning based Q-Learning method for securing a CPS. The survey focused on CPS in terms of supported techniques, domains, and attacks. The study reported in\,\cite{zhu2010scada} focused  on SCADA specific intrusion detection and prevention. The survey presented in this article focuses on behavior-based approaches for intrusion detection in CPS focusing on ML and DL techniques. Recently these approaches have gained more attention as they are relatively easier to automate than others, and are scalable and generalizable for new ICS. 

\section{Dimensions For Classifying Intrusion Detection Systems}
\label{sec:IDS}\label{sec:dimensions}

Recent progress in ML coupled with attempted and successful cyber-attacks on critical infrastructure, has sparked a wave of interest in behavior-based IDS for ICS. It is important for researchers and practitioners to understand how the  proposed approaches compare with each other and their usability in operational environments. With this as our goal, it was decided to  adopt a multi-dimensional approach to categorize the literature most of which focuses  on IDS for ICS while some on a broader class of CPS. Specifically, works surveyed in this article are placed in a 7-dimensional space where the dimensions are domain, audit material, complexity, feature selection, time series, dataset, and metrics. The use of this multi-dimensional space adds formalism to the   comparison of different works and enables a scientific discussion on their utility or non-utility in specific environments. We mote that the adoption of a multi-dimensional approach for  categorization of research has also been adopted by other researchers\,\cite{10.1145/2542049}. However, the multi-dimensional space adopted by us is richer in terms of the dimensions selected and their number. The dimensions used in the work are enumerated in Table~\ref{tab:dimensions} and described in the following subsections.

\begin{table*}
\caption{Dimensions used for categorizing literature in this survey.}
\label{tab:dimensions}
\begin{tabular}{|l|p{4in}|}
\hline
{\bf Dimension}&{\bf Description}\\
\hline
Domain&Application domain such as electric power grid and  water treatment plant.\\
Audit material&Data used in model creation\\
Complexity&Computing power needed; scalability\\
Feature selection&Selection of features to reduce overfitting \\
Time series&Modeling processes as a time series\\
Dataset&Data used; pre-collected or live; from simulation or live plant\\
Metrics& Metrics used for evaluating the effectiveness of the ML techniques used\\

\hline
\end{tabular}
\end{table*}

\subsection{Domain}
Intrusion detection for ICS has been applied in a  variety of domains, including smart utilities. Not surprisingly, most applications are in the area of energy, water and gas primarily because of the critical nature of these systems. A power grid compromised for a few seconds can trip a generator. This transfer may result in the affected load transferred to other generators and possibly initiate  a cascade of generators tripping one after the other leading to a major blackout.  The works labelled as Annon in Table\,\ref{tab:OCC},\,\ref{tab:SL(1)},\,\ref{tab:SL(2)}, and\,\ref{tab:UL1} do not specify the domain on which the proposed approach is applied,  instead they mention it as ``some CPS/ICS".

In our survey we found that the least explored ICS in smart utility  is gas. Even though a few such ICS are listed in Table\,\ref{tab:UL1},  they rely on a relatively simple gas ICS testbed at Mississippi State University (MSU)\,\cite{morris2011control}, which consists of a a minimal set of components including pressure sensor, a pump, and a solenoid valve.

\subsection{Audit Material}
Typically the data analyzed by an IDS includes   network traffic and  sensor measurements  with few IDS considering both. Since IDS were first developed for the internet and LAN networks,  most of the IDS developed for ICS also attempt to detect intrusions in the network layer using similar approaches. Typically, ICS use industrial control protocols such as Modbus\,\cite{cheung2007using}, BACnet\,\cite{sayegh2014scada}, and DNP3\,\cite{berthier2011specification}. Hence,  it is commercially viable to develop IDS for such protocols. A study reported in\,\cite{kwon2015behavior} used bits per packet, connections per second, and recent/mean interval time and count of Goose messages for this purpose. Another study reported in\,\cite{gao2010scada} used several responses against a command to detect attacks. The study in\,\cite{dussel2009cyber,hadvziosmanovic2012n} did deep packet inspection to calculate the n-gram features from the payload of the packet. The method used in this work is to constructs a feature vector that contains the count, frequency, and binary occurrence of these n-grams. The authors also argue   that n-grams are successful in detecting attacks. However, the approach proposed in\,\cite{sayegh2014scada} uses Ethernet, IP, UDP, and BACnet packet header attributes to train its IDS. The study reported in\,\cite{mantere2014module} suggests detecting attacks by using the number of live TCP, UDP \& ICMP connections, duration of terminated connections, overall network fragments pending reassembly by Bro, amount of data sent by connection responder/originator, and the number of packets sent by connection responder/originator features.

Detecting attacks in the physical process controlled by an ICS is  challenging as components, size, and functionality of each process is different from others. Such IDS have received relatively little attention and though at the time of writing this survey there seems to be  a growing trend to detect intrusion at the physical process level. IDS that model the physical process of the energy systems have used the following features to train their model:  voltage Phase angle,  voltage magnitude,  current phase angle,  current phase magnitude, zero voltage phase angle magnitude, current phase angle magnitude, the frequency of relays, frequency delta for relays, apparent impedance seen by relays, angle seen by relays, status flags for relays, snort alert status for each relay, control panel remote trip status, and their correlations\,\cite{pan2015developing,borges2014machine}. A study reported in\,\cite{david2016CRC} used the status of the pumps and valves, rate of inflow, level of the tank, and rate of change of water level for water ICS. For gas ICS,\,\cite{nader2013intrusion} uses pressure in the pipeline, pump, and solenoid status as features.

There have been few attempts in developing a hybrid approach by using both the network traffic and physical process features. A study reported in\,\cite{gao2010scada,beaver2013evaluation} used a couple of physical process features along with a few dozen network traffic features to detect attacks in gas ICS. Also, a study reported in\,\cite{yang2006anomaly} used CPU and OS usage parameters in addition to features of  network traffic  to detect attacks in a simulated CPS made up of different SUN Microsystem servers and workstations. A study  reported in\,\cite{krishnamurthy2014scalable} used  Wireshark to capture  network logs and physical stream data such as temperature and airflow. This data was then used to learn an IDS for Heating, ventilation, and air conditioning (HVACs). The above-mentioned hybrid approaches have used a single algorithm to model both the network traffic and physical processes.

\subsection{Complexity}
Based on complexity, we refer to some approaches as simple when they follow the traditional ML life cycle, i.e., derive some features, followed by some feature selection, and training a classifier. Hybrid approaches follow a more complex life cycle by either a)~transforming the input features to a transformed features space where a classifier is trained to give better performance\,\cite{sayegh2014scada}, or b)~multiple classifiers are trained separately but cooperate to arrive at a  decision\,\cite{landford2015fast,wijayasekara2014fn,sayegh2014scada}. A study reported in\,\cite{palacios2013intrusion} first used the K-means to cluster the data followed by self-organizing maps (SOM) to do the final classification. Another study reported in\,\cite{landford2015fast} learns five different SVM's and uses an ensemble of them to detect attacks. Likewise, a three-tier system for state monitoring of a CPS was proposed in\,\cite{wijayasekara2014fn}. The first tier consists of a threshold-based alarm. The minimum and maximum bound of each sensor are defined here. Anything above or below this bound triggers an alert. The second tier uses a self-organizing fuzzy logic system. The purpose of this layer is to detect anomalies. This tier learns the rules of the CPS itself.  The third tier uses an artificial neural network (ANN) to forecast the value of each sensor based on the historical data. Finally, fuzzy logic is used to raise an alarm based on outputs of tier~2 and tier~3. A study reported an anomaly-based IDS for SCADA\,\cite{palacios2013intrusion}. It extracts the time correlation between different packets using histograms, followed by Bayesian inferencing, to identify attacks. An alarm is raised if the probability of belonging to anyone of the seen categories is below a spefific threshold.

\subsection{Feature Selection}
Feature selection techniques are used to increase the accuracy and to reduce the overfitting and training time of the model; the selection  could be manual or automatic.  Feature selection techniques include Univariate Selection, Feature Importance, and Correlation Matrix with Heatmap\,\cite{featureselection}. Deep Learning techniques do not require explicit feature selection because they have  an inherent capability to select the best features for the model. A study reported in \cite{NAZIR2021102164} proposed a feature selection method based on Tabu Search and Random Forest. They used Tabu Search for searching and Random Forest as a learning algorithm for intrusion detection. 

\subsection{Time Series}
Time series data contains a well-defined time pattern consisting of a specific sequence of measurements. This property is quite useful as it helps  determine which particular algorithm, such as time series analysis or any other, would be better to apply in the ML or DL model. A study reported in\,\cite{kaburlasos2019time} used  fuzzy logic to classify the time series data of sensors in CPS. It represented the time series using the distribution of its data samples. This was done using its proposed Intervals Numbers technique. Moreover, the effectiveness of the proposed approach was tested using a benchmark classification problem.

\subsection{Dataset}
A major bottleneck in the use of supervised ML and DL techniques is the lack of attack data. The attacks on real-world systems are rare and sparse. Therefore, studies that have used actual data from CPS have resorted to simulated attacks to train and evaluate its classifiers\,\cite{linda2009neural, landford2015fast}, thus making the realism and fidelity questionable. Other works resort to validate their model on completely simulated data\,\cite{shin2010experimental,yang2006anomaly,pan2015developing,borges2014machine}. Some studies have even used the NSL-KDD99 dataset\,\cite{bay2000uci} to validate their IDS whereas, this data is a collection of simulated raw TCP dump data over nine weeks on a military local area network. It is a benchmark dataset for IDS in normal LAN traffic but not for CPS network traffic. 

A publicly available dataset is provided by a Critical Infrastructure Protection Center at Mississippi State University (MSU) \footnote[1]{https://sites.google.com/a/uah.edu/tommy-morris-uah/ics-data-sets}. Their power system dataset is a simulated smart grid data consisting of data under normal behavior, attacks, and faults. This dataset was used by\,\cite{pan2015developing,borges2014machine} for intrusion detection in ICS using ML approaches. Their water storage tank and gas pipeline dataset were developed using small scale laboratory testbed and were used by\,\cite{nader2014norms,nader2013intrusion,beaver2013evaluation} to detect intrusion in CPS using ML approaches. Their water testbed consists of a water tank having a storage capacity of 2~liters, a pump, and a level sensor. It consists of a physical process attribute for the level of the tank and the status of the pump. Apart from that, they have seventeen different network traffic and PLC status attributes. The gas pipeline dataset consists of twenty-three network attributes and PLC status attributes, and only three physical process attributes, namely pressure in the gas pipeline, solenoid, and pump status. Both of these datasets are flawed for ML research as acknowledged by the authors themselves. SWaT dataset\,\cite{goh2016dataset} is another publicly available dataset of a water treatment testbed. This testbed is an industrial scaled-down replica  of a water treatment plant. It has six stages and can produce five gallons per minute of filtered water. Data collection was done by running the plant non-stop for eleven consecutive days. For the first seven days the plant was run in a normal state while during the last four days specifically crafted attacks were launched on the plant. Therefore, this dataset contains both the normal and attack data of a real testbed. Both network and physical process data were collected for this purpose. Following  the publication of the dataset in\,\cite{goh2016dataset}, iTrust has made public several other datasets collected from the SWaT testbed\,\cite{swatDataset}. The SWaT datasets have been used in a large number of research projects including, though not limited to, \,\cite{inouYamagataChenPoskittJun,umer2017integrating, umer2020generating,junejo2020predictive,junejo2016behaviour,umer2020method}.

\begin{figure*}[tbh]
\centering
\includegraphics[width=1.0\linewidth,height=14cm]{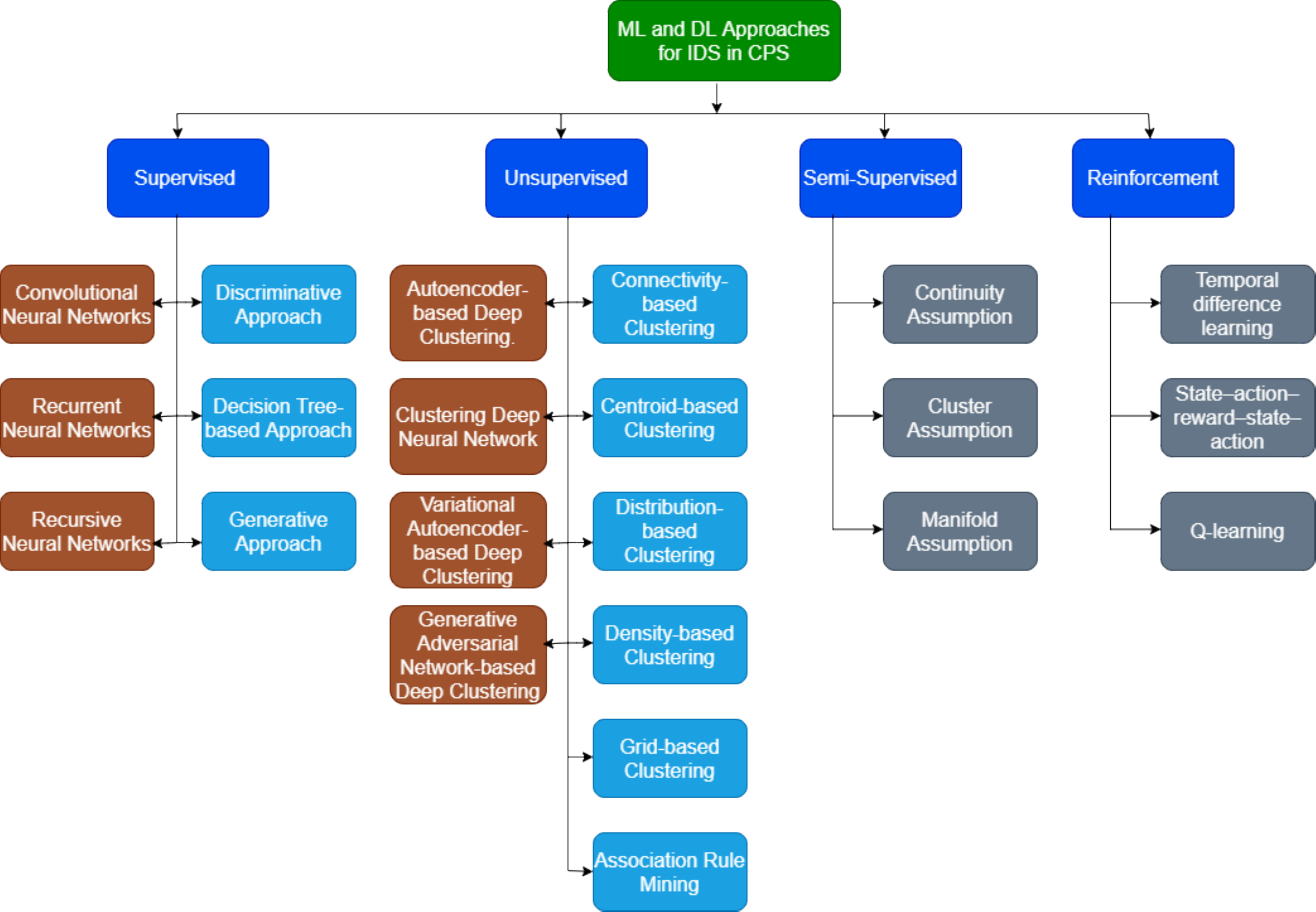}
\caption{Categorization of machine learning  approaches for detecting intrusions in Industrial Control Systems.}
\label{fig:mldl}
\end{figure*}

\begin{figure*}[tbh]
\centering
\includegraphics[width=0.75\linewidth, height=8cm]{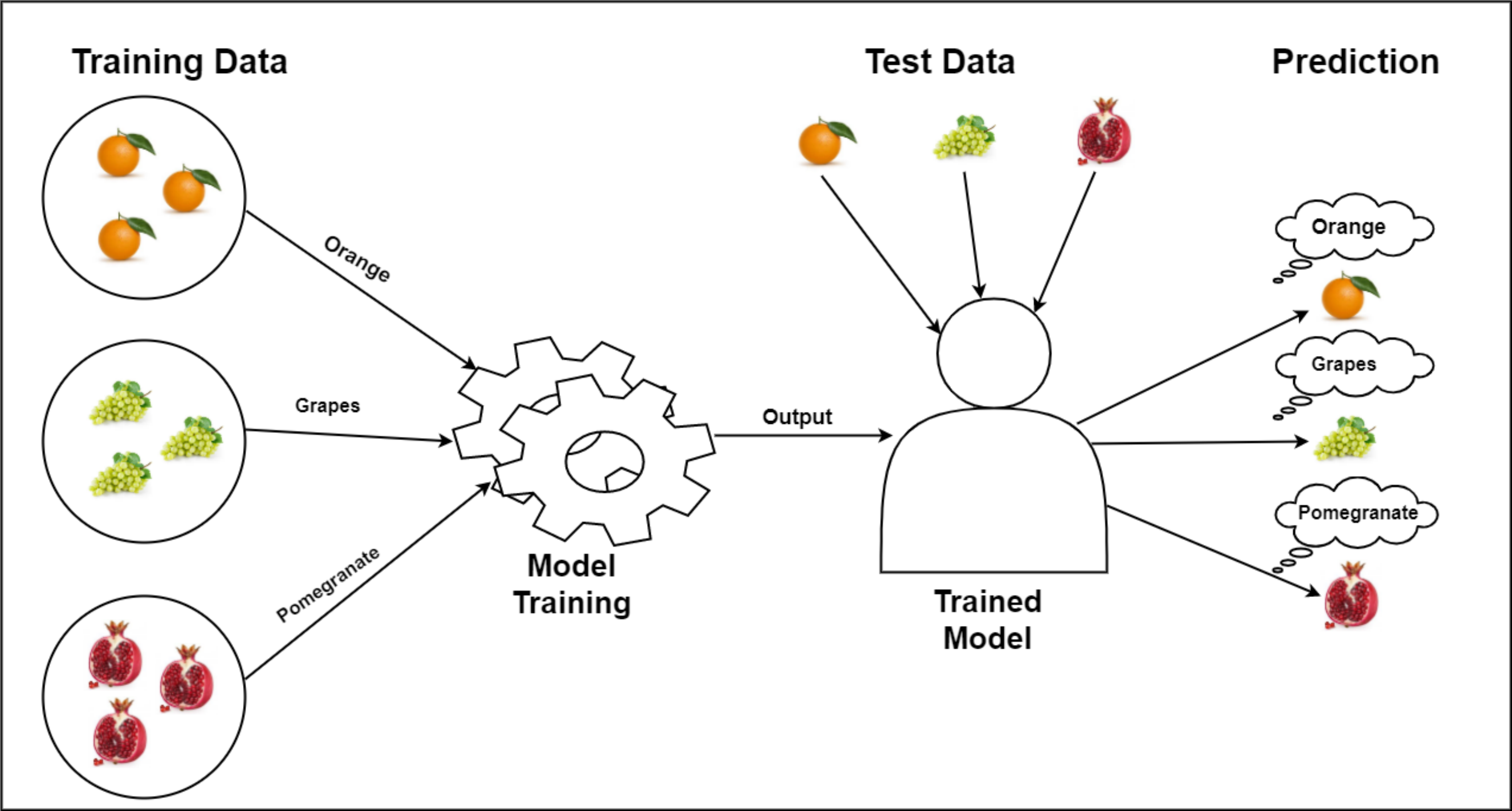}
\caption{Supervised Learning}
\label{fig:SL}
\end{figure*}

\subsection{Metrics}
\label{metrics}
Intrusion detection is a skewed class problem, also known as  class imbalance. This refers to a setting where most of the data belongs only to a single class, e.g., instances of normal behavior in an IDS dataset constitute more than 90\% of the dataset. Hence any naive classifier that labels each instance as normal will get an accuracy higher than 90\%. Therefore, accuracy is not enough to assess the performance of IDS, and yet some studies only report accuracy (or error graphs). Similarly, some studies report only the detection rate (DR), which is the same as recall. The recall alone is not enough to assess the performance of IDS, as there is a trade-off between precision and recall. A 100\% recall can always be achieved by compromising the precision of the system. 

For proper evaluation of the effectiveness of an IDS, more than one of the following metrics should be reported: accuracy, precision, recall, F-measure, receiver operating characteristic (ROC), and area under the ROC curve (AUC). Precision measures the correctness of the classifier based on the detection of an attack. A high value of precision leads to a lesser number of false positives (FP). Whereas, recall is the number of attacks detected by the classifier. A high value of recall leads to a lesser number of false negatives (FN). An ideal classifier should have high precision and recall. F-Measure helps us to combine both into a single metric, which is the harmonic mean of precision and recall. It is a more conservative measure than the arithmetic mean of the two. These measures are defined as follows.  

\begin{align*}
 Accuracy &= \frac{TP+TN}{TP + FP + TN + FN}  
 \\
 Precision &= \frac{TP}{TP + FP} \\
 Recall &= \frac{TP}{TP + FN} 
 \\
 F-Measure &= \frac{2*Precision * Recall}{Precision + Recall}
	\end{align*}
\skipnoindent where TP is the number of attacks correctly classified by the classifier, and TN the number of normal instances classified as normal.

ROC curve is a true positive rate (TPR) plotted against false positive rate (FPR) thresholded at various settings, whereas AUC is the area under this ROC. These measures are considered to be more robust for highly skewed problems\,\cite{powers2011evaluation}. The reason being that by increasing and decreasing the sensitivity of a sensor, the output of the classifier can often be tweaked to make it more (or less) conservative thus achieving a trade-off between FP and FN. The AUC measure allows selecting possibly optimal models by evaluating the performance of the classifier by varying the threshold that decides whether the instance is an attack or not. Unfortunately, few researchers\,\cite{david2016CRC, kosmanos2020novel,patel2017nifty,whalen2014model,o2016distributed} have used this measure for evaluating their respective classifiers. 

The time to detect an attack and the percentage of the time the attack remains detected should also be used as an evaluation metric. It is likely not of any value when an  attack is detected once it has already damaged a physical component or the attack is detected intermittently by turning on and off the alarm after every few seconds leading to confusion. Few studies report these measures \cite{landford2015fast, umer2020generating,david2016CRC}. Among these,\,\cite{landford2015fast} reported the latency. They define latency as the number of cycles after the spoof begins but before the classifier correctly identifies a string of 30 consecutive cycles as spoofed. Whereas\,\cite{david2016CRC, umer2020generating} reported the time to detect an attack. While\,\cite{david2016CRC} also reported the percentage of  time the attack was detected during its course.

\section{Machine Learning Approaches For Intrusion Detection}
\label{sec:MLApproaches}

As shown in Figure\,\ref{fig:mldl}, ML and DL techniques can be classified into four major categories, i.e. Supervised Learning, Unsupervised Learning, Semi-Supervised Learning, and Reinforcement Learning. Most of the intrusion detection work available in the literature is related to the first two areas while  limited work is available in the last two areas. The difference between the first three approaches lies in whether  or not the training data used is labeled. An unsupervised approach does not require labeled data, relying solely on the normal behavior of the ICS. A supervised approach requires training data under both normal and abnormal (attack) behavior. The semi-supervised approach makes use of both, relying on the assumption that labeled training data is scarce and rare whereas unlabeled training data is plenty and easily available. All  areas mentioned in  Figure\,\ref{fig:mldl} are discussed in subsequent sections. 

Each  approach  mentioned above has its pros and cons. Unsupervised learning does not require labeled training data, therefore, the dependency on attack data gets eliminated making it capable of detecting zero-day attacks. However, it usually produces high false alarms\,\cite{nader2014norms,nader2014mahalanobis}. While the supervised learning algorithms are more robust in terms of attack detection, they require labeled data, i.e., both normal and attack data. Given only a few instances of attacks, the supervised approach is capable of detecting other instances of attacks as well. The study reported in\,\cite{david2016CRC} showed that their best classifiers produce almost no FPs and achieved high precision and recall. These approaches do not have any guarantee in detecting the zero-day attacks.

Another set of promising approaches that have not been explored for IDS in ICS are one-shot learning\,\cite{wu2012one,krishnan2015conditional} and zero-shot learning\,\cite{romera2015embarrassingly,socher2013zero}. One-shot learning refers to a scenario where only one instance of each attack type is available in the labeled training data. Whereas zero-shot is a more challenging approach in which few instances of some attacks are available in the labeled training data. The attack type that does not have any instance in the training data represents zero-day attacks. Thus, the performance of this type of learning is based on detecting the zero-day attacks while leveraging the information provided by the known attacks. This represents a more practical approach for an ICS as it would generate fewer FPs than unsupervised approaches and at the same time detect zero-day attacks while leveraging on some known attacks that can be safely carried out on the ICS in a controlled environment. We believe that zero-shot learning is a promising approach for IDS in ICS because it achieves a good compromise between the supervised and unsupervised approaches.

\section{Supervised Learning}
\label{sec:supervisedLearning}
Supervised Learning (SL) requires labeled training data as described in Figure\,\ref{fig:SL}. For each instance of the training dataset, SL uses 'n'  features from feature vector 'X', i.e., [x\textsubscript{1}, x\textsubscript{2} .....,x\textsubscript{n}] to learn the class variable 'Y' against each instance of the dataset. The relationship between 'X' and 'Y' is captured in the equation $\text{Y} = f(\text{X})$ where $f$ is learned from data.

There are mainly two types of SL techniques: classification and regression. In classification, the class variable is discrete while for regression problems it is continuous. IDS are typically modeled as classification problems where the class variable can contain both single and multiple classes. If the class variable contains only a single class then it is referred to as a One-Class Classification (OCC) problem. OCC-based IDS research for ICS is summarized in Table\,\ref{tab:OCC} while the work related to multiple classes is summarized in Table\,\ref{tab:SL(1)} and \ref{tab:SL(2)}


\subsection{Supervised Learning Approaches}
Certain behavior-based approaches have used conventional statistical techniques\,\cite{yang2006anomaly,kwon2015behavior}. These approaches use traditional statistical techniques such as  mean and standard deviation on sensor measurements. These techniques are not completely automatic due to their parametric nature. It is difficult to produce statistical tests for a deeply interdependent and large number of sensors and actuators as doing so may lead to unacceptable FPs. ML and DL are considered as non-parametric approaches. They are more automatable and diverse in terms of different techniques employed while using them. In this survey we  have grouped the ML approaches as discriminative, generative, and tree-based, with details of each given below.

\newcolumntype{C}[1]{>{\centering\let\newline\\\arraybackslash\hspace{0pt}}m{#1}}
\begin{table*}[tbh]
\caption{Summary of OCC-based Intrusion Detection work in ICS using Supervised Learning techniques}
\label{tab:OCC}
\resizebox{18.15cm}{5.25cm}{
\begin{tabular}{C{1cm}C{1.5cm}C{1.5cm}C{1.75cm}C{2cm}C{1.5cm}C{1cm}C{2cm}C{1.5cm}C{1.4cm}C{3cm}}
\hline
\textbf{Work} & \textbf{Domain} & \textbf{Audit Material} & \textbf{Complexity} & \textbf{Algorithms} & \textbf{Feature Selection} & \textbf{Time Series} & \textbf{Dataset} & \textbf{Data Type} & \textbf{Data Available} & \textbf{Metrics} \\ \hline
\hline
\cite{kreimel2017anomaly} & Conveyor Belt System & Physical & Simple & k-NN, and NB & Yes & Yes & Annon & Actual & No & Confidence, Accuracy \\ 
\cite{krishnamurthy2019anomaly} & Chemical Plant & Physical & Simple & OCSVM & Yes & Yes & HITL \ignoretext{(Hardware in the loop)}  & Actual & Yes & Accuracy, Precision, Recall, and F1 score \\ 
\cite{wang2016malicious} & Annon & Physical & Simple & OCSVM & Yes & Annon & Annon & Actual & No & FPR, FNR \\ 
\cite{nader2014norms} & Gas, and Water & Physical & Simple & SVDD, and KPCA & No & Yes & \ignoretext{Mississippi State University,and University of California } MSU, and UCI & Actual & Yes & Accuracy \\ 
\cite{zizzo2019adversarial} & Water & Physical & Simple & LSTM & Yes & Yes & SWaT & Actual & Yes & Accuracy \\ 
\cite{li2020non} & Industrial Demonstrator & Physical & Simple & OCSVM, DINA & No & Yes & Industrial demonstrator\ignoretext{ (Genesis demonstrator)}, and Wind Turbines & Actual, and Simulated & No & TPR, TNR, F1 Score, and Balanced Accuracy \\ 
\cite{demertzis2019gryphon} & Water, Gas, and Energy & Nerwork & Simple & ESNN, SOCCADF, OCC-SVM, OCC-CD/CPE & No & Yes & Water, Gas, and Electric & Actual & Yes & TPR, TNR, TA, Precision, Recall, and F1-score \\ 
\cite{zhang2012diagnosing} & Annon & Network & Hybrid & SVM & No & Yes & Annon & Actual & No & DR, and IR \\ 
\cite{wang2018distributed} & Energy & Physical & Hybrid & DAE, OCSVM, AdaBoost + C4.5, XGBoost, MLP, SVM, k-NN. & Yes & Yes & Annon & Simulated & No & Accuracy, Precision, Recall, and F1 score \\ 
\cite{foroutan2017detection} & Energy & Physical & Simple & GDLM, SVM, MLP, and PCA & No & Yes & Annon & Simulated & No & F1 Score \\ \hline
\end{tabular}
}

\end{table*}

\subsubsection{Discriminative Approaches}
Support Vector Machines (SVM) are linear classifiers,  non-probabilistic, and perform binary classification. When using SVM, the data points are projected to a higher dimensional feature space. Then, a hyperplane is learned to distinguish the data points of the two classes. The goal of learning a hyperplane is to enlarge the difference between the closest data points of the classes and thereby provide stronger generalization on the unseen data. This property of SVM makes it robust for classification problems  including IDS\,\cite{ahmad2014enhancing}.

Artificial Neural Networks (ANN) is a class of algorithms that attempt to mimic the learning process of biological neural networks. ANNs are capable of estimating the functions that are dependent on a large number of inputs. There are multiple layers in this network including input, output, and one or more hidden layers. It trains the model to learn the non-linear decision boundaries to segregate the classes. ANNs have also been used for IDS\,\cite{al2015network}.

Instance-based learning algorithms (IBK) do not work on generalization as compared to SVM and ANN. Instead, they compute the distance of every new instance with all the available instances in the training dataset. A decision is taken based on all the computed distances. That is why IBK is also referred to as a lazy learning algorithm. It has been used in\,\cite{palacios2013intrusion, muda2011intrusion, kumar2013k} for IDS. The Non-Nested Generalized Exemplars (NNGE) also belong to this class of algorithms\,\cite{panda2009ensembling}. It was applied to detect network intrusions in KDDCup 1999 dataset\,\cite{bay2000uci}.

Artificial immune systems try to mimic the complex vertebrate immune system\,\cite{zhang2011artificial}. They are intelligent and robust computing systems. Fuzzy rules were developed in\,\cite{wijayasekara2014fn} to express the normal behavior of the system. This was done using Fuzzy-Neural Data Fusion Engine (FN-DFE). Later these fuzzy rules were used for anomaly detection by comparing it with previously described rules of the system. Moreover, a classifier based on neural networks was used to make the concluding decision based on these anomalies. 

Multinomial Logistic Regression (LR) is  comparable to linear regression and serves as an alternative to Linear Discriminant Analysis (LDA). However, they both have different underlying assumptions. LR assumes Bernoulli distribution while linear regression assumes Gaussian distribution. Moreover, LR uses the logistic function for prediction. The so predicted values are the probabilities calculated using the logistic function and measure the relationship between the dependent and the independent variable(s). Here, the dependent variable is categorical. Its performance can be improved by using a large number of features. However, it is not as successful in IDS\,\cite{tsai2009intrusion}.

\begin{table*}[tbh]
\caption{Summary of Multiclass-based Intrusion Detection in CPS using Supervised Learning technique (1 of 2)}
\label{tab:SL(1)}
\resizebox{18.15cm}{10.2cm}{
\begin{tabular}{C{1cm}C{1.5cm}C{1.5cm}C{1.75cm}C{2cm}C{1.5cm}C{1cm}C{1.5cm}C{1.5cm}C{1.5cm}C{3cm}}
\hline
\textbf{Work} & \textbf{Domain} & \textbf{Audit Material} & \textbf{Complexity} & \textbf{Algorithms} & \textbf{Feature Selection} & \textbf{Time Series} & \textbf{Dataset} & \textbf{Data Type} & \textbf{Data Available} & \textbf{Metrics} \\ \hline \hline
\cite{wang2017novel} & Energy & Physical & Simple & MSA, SVM, and ANN & No & Yes & PMU & Actual, and Simulated & No & Accuracy \\ 
\cite{li2018towards} & Healthcare & Physical & Simple & k-NN, NN, SVM, DT, NB, and ZeroR. & No & Yes & Annon & Actual, and Simulated & No & Accuracy, Precision, Recall,and F1 Score \\ 
\cite{elgendi2019protecting} & Water & Network, and Physical & Hybrid & SVM, and SMC & Yes & Yes & Annon & Actual & No & Accuracy, Sensitivity, and Specificity \\ 
\cite{anthi2019supervised} & Smart Home & Network & Simple & NB, BN, J48, Zero R, OneR, Logistic, SVM, MLP, and RF & Yes & Yes & Annon & Actual & No & F1 Score \\ 
\cite{soltan2019line} & Energy & Physical & SImple & BR with ARD & No & Yes & Annon & Simulated & No & FP, FN, and PT \\ 
\cite{yan2019attack} & Gas, and Energy & Physical & Simple & ELM & Yes & Yes & Annon & Actual, and Simulated & No & ROC, TPR, and FPR \\ 
\cite{ahmed2018noiseprint} & Water & Network, and Physical & Simple & SVM & Yes & Yes & SWaT, and WADI & Actual & Yes & Accuracy \\ 
\cite{amrouch2017emerging} & Annon & Physical & SImple & CNN & Yes & Yes & Annon & Actual & No & Accuracy \\ 
\cite{ghaeini2019zero} & Water & Network, and Physical & Simple & RF, NBTree, LMT, J48, PART, MLP, HTree, LogF, and SVM. & Yes & Yes & SWaT & Actual & Yes & Precision, and Sensitivity \\ 
\cite{sokolov2019applying} & Water & Physical & Simple & NN & Yes & Yes & SWaT & Actual & Yes & Accuracy, Precision, Recall, and F1 score \\ 
\cite{kosmanos2020novel} & Electric Vehicles & Network, and Physical & Hybrid & RF, and k-NN & No & Yes & Annon & SImulated & No & Accuracy, DR, ROC, and AUC \\ 
\cite{ariharan2019machine} & Annon & Physical & Hybrid & LSTM, NN, SVC, and SVM & Yes & Yes & Annon & SImulated & No & Probability of detection \\ 
\cite{shenfield2018intelligent} & Annon & Network & Simple & ANN & No & Yes & Annon & Actual & No & Accuracy, Precision, Sensitivity, and ROC \\ 
\cite{kumara2018automated} & Cloud & Physical & Simple & LR, RF, NB, RT, SMO, and J48 & Yes & Yes & Annon & Actual & No & TPR, TNR, F1 score,and Accuracy \\ 
\cite{patel2017nifty} & Energy & Network & Simple & SVM & No & Yes & Annon & Actual & No & Accuracy, and AUC \\ 
\cite{feng2020efficient} & Drones & Physical & Simple & GA, XGBoost, and SVM & No & Yes & Annon & Actual & No & Precision \\ 
\cite{stockman2019detecting} & Energy & Physical & Simple & SVM, k-NN, RF, and CNN & Yes & Yes & Annon & Actual & No & Accuracy \\ 
\cite{kozik2018scalable} & Cloud & Network & Simple & ELM & Yes & Yes & CTU & Actual & Yes & TPR, FPR, TNR, FNR, Precision, Accuracy, ER, F1 score, MC \\ 
\cite{sharma2018hybrid} & VANETs & Network & Simple & SVM & Yes & Yes & Annon & SImulated & No & DE, FPR, DT and CH Load 
 \\ 
 \cite{ieracitano2019novel} & Annon & Network & Simple & AE, LSTM, MLP, SVM, LDA and QDA & Yes & Yes & NSL-KDD & Actual & Yes & Precision, Recall, F1 score, and Accuracy\\
\cite{raman2019efficient} & Annon & Nework & Simple & BN, NB, MLPNN, J48, and SVM & Yes & Yes & NSL-KDD CUP, and UNSW-NB15 & Actual & Yes & DR, FAR, and Accuracy \\ \hline

\end{tabular}
}

\end{table*}

\begin{table*}[tbh]
\caption{Summary of Multiclass-based Intrusion Detection in CPS using Supervised Learning technique (2 of 2)}
\label{tab:SL(2)}
\resizebox{18.15cm}{10.2cm}{
\begin{tabular}{C{1cm}C{1.5cm}C{1.5cm}C{1.75cm}C{2cm}C{1.5cm}C{1cm}C{1.5cm}C{1.5cm}C{1.5cm}C{3cm}}
\hline
\textbf{Work} & \textbf{Domain} & \textbf{Audit Material} & \textbf{Complexity} & \textbf{Algorithms} & \textbf{Feature Selection} & \textbf{Time Series} & \textbf{Dataset} & \textbf{Data Type} & \textbf{Data Available} & \textbf{Metrics} \\ \hline \hline
\cite{ghanem2019new} & Annon & Network & Simple & MLP & Yes & Yes & KDD Cup 99, NSL-KDD, SCX2012, and UNSW-NB15 & Actual & Yes & DR, FAR, and AR \\ 
\cite{haghnegahdar2019whale} & Energy & Physical & Simple & RF, OneR, JRip, Adaboost + JRip,SVM, and NN & Yes & Yes & MSU, and ORNL & Actual & Yes & Accuracy, Precision, Recall, and F1 score \\ 
\cite{li2020aquasee} & Supercomputer/ Water & Physical & Simple & LSTM & No & Yes & Tianhe-1A & Actual & Yes & RMSE, and Accuracy \\ 
\cite{rathore2018multi} & Healthcare & Physical & Simple & MLP, and SVM & Yes & No & ECG-ID & Actual & Yes & Accuracy, Precision, Recall, and F1 score \\ 
\cite{sheikhan2014flow} & Annon & Network & Simple & MLP, MGSA, PSO, and EBP & No & Yes & Intrusion Detection dataset & Actual & Yes & CCR, ER, MR, and FAR \\ 
\cite{otoum2019feasibility} & Annon & Network & Simple & ASCH-IDS, and RBC-IDS & Yes & No & KDD’99 Dataset & Simulated & Yes & AR, FNR, DR, ROC , and F1 score \\ 
\cite{zhang2018response} & Energy & Network, and Physical & Hybrid & BPNN, and ELM & Yes & Yes & Annon & SImulated & No & Error / Hz \\ 
\cite{loukas2017cloud} & Vehicle & Network, and Physical & SImple & RNN, MLP, LR, DT(5.0), RF, and SVM & Yes & Yes & Annon & Actual & No & Accuracy \\ 
\cite{palacios2013intrusion} & Annon & Network & Hybrid & k-means-SOM & No & No & KDDCup1999 & Actual & Yes & FPR, TPR, and DR \\ 
\cite{zhang2011distributed} & Energy & Network & Simple & SVM, and AIS & No & No & KDDCup1999 & Simulated & Yes & FPR, FNR, and No. of Detections \\ 
\cite{landford2015fast} & Energy & Physical & Hybrid & Ensemble of SVMs & No & Yes & Bonneville Power Administration & Actual & No & Recall, Precision, F1 score, and Latency \\ 
\cite{pan2015developing} & Energy & Physical & Hybrid & CPM & No & No & MSU Power & Simulated & Yes & Accuracy, and FPR \\ 
\cite{borges2014machine} & Energy & Physical & Simple & NB, OneR, Nnge, Jripper, RF, SVM, and Adaboost & No & No & MSU Power & Simulated & Yes & F1 score \\ 
\cite{wijayasekara2014fn} & Energy & Physical & Hybrid & Fuzzy-Neural Data Fusion Engine & No & No & Idaho National Labs energy sys. model & Actual & No & Error Graphs \\ 
\cite{beaver2013evaluation} & Gas & Hybrid & Simple & NB, OneR, Nnge, RF, SVM, and J48 & No & No & MSU & Actual & Yes & Precision, and Recall \\ 
\cite{gao2010scada} & Water & Hybrid & Simple & NN & No & No & MSU & Actual & No & Accuracy, FP, and FN \\ 
\cite{david2016CRC} & Water & Physical & Simple & RF, SVM, NN, J48, BN, NB, BFTree, BayesLR, LR, and IBK & No & No & SWaT & Actual & No & Accuracy, AUC, Precision, Recall, and F1 score \\ 
\cite{lin2018tabor} & Water & Physical & SImple & RTI+, and BN & Yes & Yes & SWaT & Actual & Yes & CP, and PS \\ \hline
\end{tabular}
}

\end{table*}

\subsubsection{Decision Tree-Based Approaches}
This class also belongs to the discriminative-based approaches but are classified separately due to the existence of  distinctive features. This class has been  popular  among ML researchers.  The decisions in this class can be easily translated into an IF-ELSE structure using logical connectives like OR, AND, etc. These decisions (rules) are impulsive and easy to understand. These decisions follow a tree-like structure having nodes from the top (root) to bottom (leaves). Here the internal nodes can be considered as a test on a feature (attribute). The branches represent the result of the test while leaves represent the labels of the class. Every new record is assigned a label by traversing the tree from the top to the bottom. The selection of attributes as different nodes of the tree is determined based on the information provided by that attribute. In ID3 and J48, this information is calculated through information gain\,\cite{yuxun2010improved, quinlan1996improved}. Information gain is the anticipated reduction in entropy by segregating the examples of datasets based on an attribute. Overfitting can be avoided by proper pruning of the tree. The traversal order of tree is an important factor in this class of algorithm. For example, J48 and Best First Tree (BFTree) are similar to each other. However, BFTree prefers the best node rather than depth-first order. This is a useful approach to prune the trees for avoiding overfitting. One Rule (OneR) is another algorithm of this class. It has one rule for each predictor of the class. The rule with smaller error is selected as "One Rule". 

Random Forest (RF) follows an ensemble learning approach\,\cite{breiman2001random}. It trains multiple decision trees based on the random subset of features. The majority vote from different decision trees for an instance is selected as the class of that instance. Due to the random selection of features, RF shows different accuracies in every iteration even for the same set of parameters. It is robust in terms of overfitting as compared to the other decision trees. The decision tree algorithms have enjoyed success in IDS at network level\,\cite{sahu2015network,hasan2014support}. An ensemble method (AdaBoost) was used in\,\cite{moustafa2018ensemble} for intrusion detection in the network traffic of IoT devices. While IoT devices are playing a vital role in providing comfort to daily routine tasks,   they generally have a weak security mechanism. The proposed study used a hybrid approach by using multiple classifiers to detect anomalies. It used Decision Tree, Naive Bayes, and Artificial Neural Network for this purpose. Although the performance of the model is acceptable it suffered from false positives. Also, the ensemble method has more processing time than Decision Tree, Naive Bayes, and Artificial Neural Networks.

\subsubsection{Generative Approaches}
Generative approaches include Bayesian Classifiers, also referred to as probabilistic classifiers. They predict the class based on the probabilities of any object belonging to a certain class. Bayesian Networks (BayesNet) and Naive Bayes (NB) are two popular Bayesian classifiers used in IDS\,\cite{koc2012network,xiao2014bayesian}. The attributes in the NB classifier do not affect each other given the value of the class. They are scalable and the parameter requirement is linear in terms of the number of features. It is suitable for high dimensional data with good generalization over the unseen data. The model is learned in a single iteration over the train data. 

BayesNet are directed acyclic graph\,\cite{friedman1997bayesian} that represent the set of random variables along with their conditional dependencies. The dependency between the variables can be eliminated by connecting them by an edge. In the real-world, the attributes of a dataset are likely dependent on each other thus making BayesNet approach better than NB, therefore the BayesNet approach is better than NB on a small number of features. BayesLR is a Bayesian variant of LR. It uses Laplace prior to escape from  overfitting, and hence is more robust than LR for large feature space\,\cite{genkin2007large}. 

Due to their simplicity, performance, and low computational complexity,  Bayesian classifiers are commonly used to solve real-world problems.  A study reported in\,\cite{lin2018tabor} used the Bayesian networks and RTI+ (Radio Tomographic Imaging) to model the normal behavior of a system and for  anomaly detection. Naive Bayes, together  with several ML algorithms, was used in\,\cite{kumara2018automated} to protect the hypervisor or the monitor of a virtual machine. The proposed architecture is composed of executable file extractor, online malware detector, and offline malware classifier. Offline malware classification was accomplished using ML algorithms applied to benign and malicious data.

\subsection{Deep Learning Based Supervised Learning Approaches}
\label{DLSL}
Deep learning is an extension of ML which focuses on the Artificial Neural Network (ANN). It does not require a complex set of features to be manually engineered by humans, instead they aim to learn these features themselves. This makes them a promising approach for ICS as each ICS has different physical dynamics. Moreover, deep learning is capable of dealing with high-velocity data\,\cite{koroniotis2019forensics}, thus making it desirable for ICS. However, computational complexities associated with deep learning\,\cite{chen2019deep} make it difficult.   

\subsubsection{Convolutional Neural Networks}
Convolutional Neural Network (CNN) is a type of deep neural network. It normally works on visual images. CNN is a variant of Multi-Layer Perceptron (MLP). One of the important properties of MLP is Fully Connectedness, in which every single neuron in one layer has connectivity with all neurons in the next layer. This may create the problem of overfitting. To reduce this overfitting, regularization methods are applied. Regularization methods include adjustments of weights to configure the loss function. It exploits the underlying hierarchical pattern of data to get complex patterns from relatively small and simple patterns. CNN has been used in a classification model for different PLC programs using phasor measurement unit (PMU) data generated during the execution of different PLC programs\cite{stockman2019detecting}. Later it was used for the anomaly detection in the PMU data. CNN was used to detect keystroke using sensor data of nearby mobile phone\cite{giallanza2019keyboard}. They classified keystrokes using a real-world dataset of 20 users. CNN was used for anomaly detection using thermal side channels \cite{amrouch2017emerging}. Thermal images were captured on a predefine time window then these images were input in CNN to detect anomalies using the information of predefined actual active time.

\subsubsection{Recurrent and Recursive Neural Networks}
Recurrent Neural Networks (RNN) is a class of ANN. Its edges input the next time step instead of the next layer of the current time step\,\cite{rnn}. RNNs refer to two classes of networks, namely,  finite impulse and infinite impulse networks. Both classes have temporal dynamic behavior\,\cite{miljanovic2012comparative} and  could have additional stored states where storage would be controlled by the neural network. Recursive Neural Networks are also a type of deep neural network. They apply the same set of weights recursively over a structured input sequence. Therefore, they give structured predictions over variable-sized input sequences. Compared to RNN, Recursive Neural Networks work hierarchically on the input sequence\,\cite{rnn}.

Study reported in\,\cite{loukas2017cloud} makes use of RNN to protect vehicles from cyber-attacks. All computations were performed on the cloud. Long short-term memory (LSTM) networks are a type of RNN. They are effective in speech recognition and showed remarkable performance in speech applications\,\cite{fernandez2007application}. Though machine learning is being used for intrusion detection, at the same time Adversarial machine learning is being used to counter it. For example, in\,\cite{zizzo2019adversarial} LSTM was used to train the model on the normal data from a water treatment plant and its performance tested on attack data. Further, the adaptive attacks were performed to deteriorate the performance of the classifier.

State of the art classifiers, including LSTM, were applied on the NSL-KDD dataset to classify the data into different classes for an IDS\,\cite{ieracitano2019novel}. A layered architecture using different ML techniques for proactive fault management by predicting sensor values at different stages was proposed in\,\cite{ariharan2019machine}. A hybrid machine learning approach was used to detect anomalies in a simulated IoT environment. Four algorithms, including LSTM, single-layer neural network, SVC, and SVM were used in the  anomaly detector. A three-stage layered architecture was proposed for this purpose and served as the quorum for the final decision of the model. 

Accurate prediction of faults in supercomputers can be used to overcome financial losses. The historical chilled water data was used in\,\cite{li2020aquasee} to predict the load of a supercomputer using LSTM. Later, a Z-score model of predicted values was used to identify the anomalies. 5G Networks has proposed a formidable challenge to the security of data, although several anomaly detection mechanisms exist but the emergence of 5G Networks has posed a significant threat due to its high velocity and veracity of data. Deep learning methods were used in\,\cite{maimo2018self} for anomaly detection in 5G networks; this was done hierarchically. At the initial level, Deep Belief Network (DBN), or a Stacked AutoEncoder (SAE), was selected to detect the anomaly. At this level, the primary intention was to classify the anomalous data at the high velocity to cope with higher velocity data of 5G Networks and therefore accuracy was not the major concern in this phase. In the subsequent  phase, the output of DBN was used by LSTM to recognize the temporal patterns of cyber-attacks.

\section{Unsupervised Learning}
\label{sec:unsupervisedLearning}
\begin{figure*}[tbh]
\centering
\includegraphics[width=0.65\linewidth, height=5.5cm]{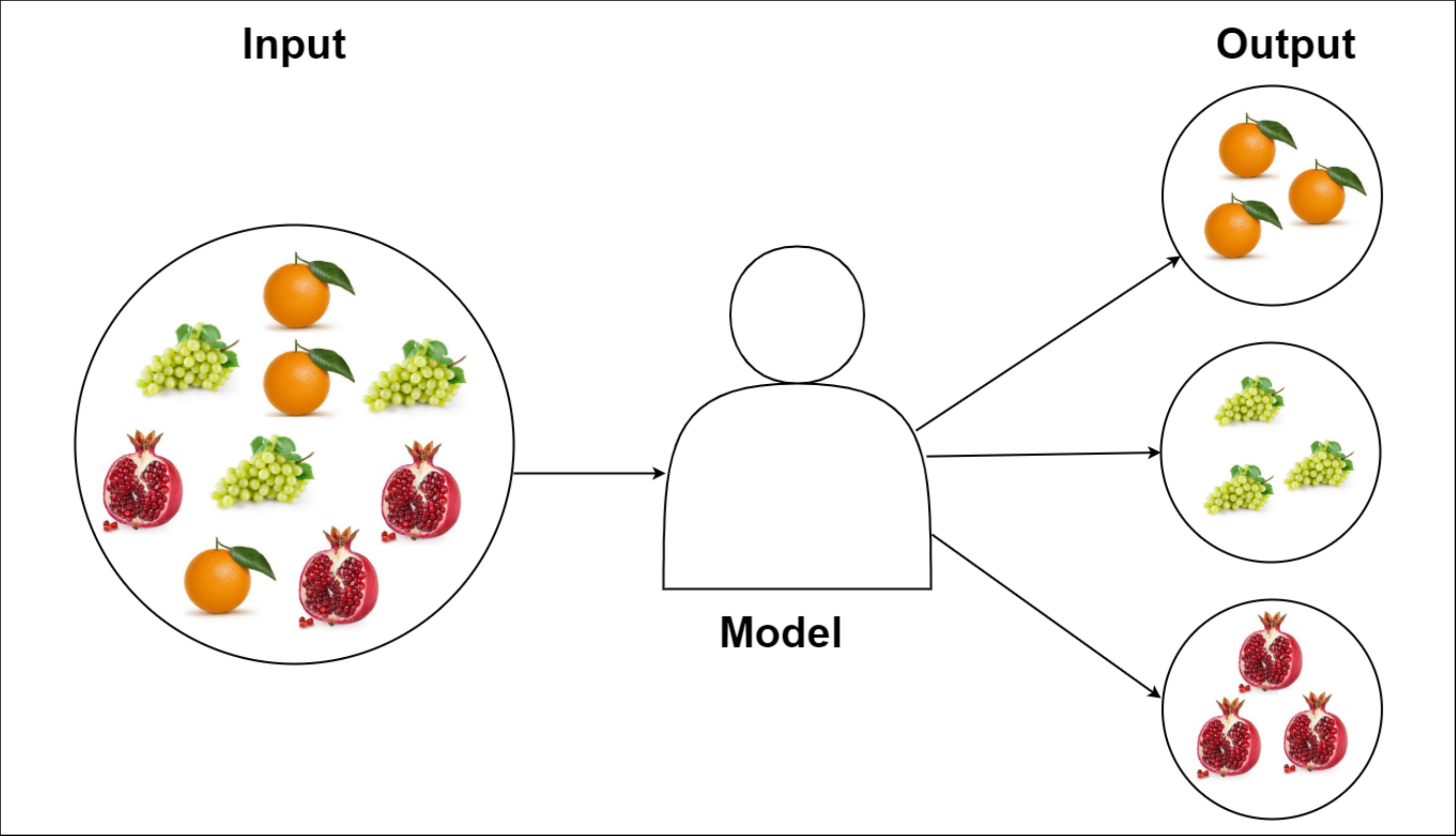}
\caption{Unsupervised Learning}
\label{fig:UL}
\end{figure*}

Unsupervised Learning (UL)  uses features from feature vector 'X', but there is no corresponding class variable as described in Figure\,\ref{fig:UL}.  Two types of UL techniques  are popular, namely,  clustering and Association Rule Mining (ARM). In clustering, different clusters are formed based on a  set of feature values while in ARM, rules are extracted  based on the support and confidence. The overall work available in literature using UL is summarized in table\,\ref{tab:UL1}.

\subsection{Unsupervised Learning Approaches}
\subsubsection{Connectivity-based clustering}
Connectivity-based clustering is also known as hierarchical clustering, i.e. a hierarchy of clusters is formed. The method  can be divided into the following two categories\,\cite{rokach2005clustering}: agglomerative clustering and divisive clustering.  Agglomerative clustering  proceeds in a bottom-up manner. Here different instances form a cluster with the nearest one at each level of the hierarchy and ultimately forms a single cluster at the top. Divisive clustering
proceeds in a top-down manner. In the beginning, a single large cluster is formed which is subsequently divided into  smaller clusters at each level of the hierarchy.

\subsubsection{Centroid-based clustering}
Centroid-based clustering is the most commonly used clustering technique.  It works on the concept of the central vector.  This central vector does not need to be a member of the dataset. Different clusters are formed based on the central vector. Each instance from the dataset is assigned to each cluster based on its distance to that cluster. K-means is a commonly use centroid-based clustering technique, where $k$ is the fixed number of clusters formed. K-means clustering was used in\,\cite{bhattacharjee2018towards} for the classification of compromised meters in the Advanced Metering Infrastructure (AMI). AMIs are highly vulnerable to false data injection attacks and can be compromised by adversaries to send false data regarding power consumption. In addition to electricity theft,  such attacks may also affect the load balancing and other critical  functions in a power grid. A consensus correction scheme was introduced in\,\cite{bhattacharjee2018towards} to detect anomaly using the ratio of harmonic to the arithmetic mean. Compromised meter classification was done using k-means clustering. The GRYPHON model is proposed in\,\cite{demertzis2019gryphon} for anomaly detection in critical infrastructure using evolving spiking neural networks, fuzzy logic, and clustering techniques. It uses fuzzy c-means clustering by assigning random values to cluster centers and subsequently assigns data points to all clusters using the Euclidean distance. 

To overcome the vulnerabilities of PLCs, a mechanism to augment PLCs with AES - 256 Encryption and Decryption was proposed in\,\cite{alves2018embedding}. Further, k-means clustering and Local Outlier Factor (LOF) was used to propose an ML-based intrusion prevention system against three categories of cyber-attacks including interception, injection, and denial of Service. A study reported in \cite{liu2014practical} used the Channel State Information (CSI) to identify the malicious user in the network. For this purpose, k-means clustering was used to differentiate malicious and legitimate users. Further, this information was used to create an Attack Resilient Profile Builder and Profile Matching Authenticator. Profile Matching was done using SVM. 

\subsubsection{Distribution-based clustering}
Distribution-based clustering is a statistical technique. It works on the principle that if objects belong to the same distribution, then they must be assigned to the same clusters. The technique usually suffers from overfitting unless constraints are applied to the complexity of the model.  Gaussian mixture model was applied in\,\cite{foroutan2017detection} to detect false data injection attacks. A mixture Gaussian distribution (MGD) was used to learn the model over normal data.Based on the parameters of this distribution, any upcoming transaction is classified as normal or anomalous. In addition, Principal Component Analysis (PCA), which is an unsupervised machine learning technique, was used for dimensionality reduction. The performance of proposed method was compared with one-class classification (OCC) by using only the normal data. OCC creates a decision boundary on the normal data so that any new transaction on the dataset could be detected whether it is an anomaly or not. The proposed method was also compared with Support Vector Machine (SVM) and Multi-Layer Perceptron (MLP). Overall, the study reported has a good F1 score. The proposed approach has better time complexity than when using SVM and MLP while lower than OCC on training data. It performed better than all of the aforementioned approaches on test data.

\subsubsection{Density-based clustering}
Density-based clustering works on the principle that higher density data areas need to be separated from the rest of the data. Doing so helps in removing  noise and in the creation of a decision boundary. Density-based spatial clustering of applications with noise (DBSCAN) is a well-known density-based clustering technique\,\cite{ester1996density}. It works on the principle of ``Density-reachability" using a distance threshold. DBSCAN was used in\,\cite{ccelik2011anomaly} for anomaly detection in temperature data. Its performance was compared against statistical approaches and several advantages observed in anomaly detection. Likewise, DBSCAN-OD, a variant of DBSCAN for outlier detection, was proposed in\,\cite{abid2017outlier} for applications with noise. It was able to detect outliers with an accuracy of 99\% in simulations.

\subsubsection{Grid-based clustering}
Grid-based clustering is used in multi-dimensional datasets \cite{aggarwal2014data}. A grid structure is created in this technique and clusters are formed by traversing   each cell in the grid  based on the threshold density. Grid-based clustering was used in\,\cite{zhong2011grid} for anomaly detection. They evaluated the system using the Kyoto2006+ and the KDD Cup 1999 datasets. False Positive rate of the proposed algorithm was better than the Song based K-means\,\cite{song2008clustering}, Song based One-Class SVM\,\cite{song2009unsupervised}, Y-means\,\cite{ymeans}, k-means\,\cite{macqueen1967some}, and Li\,\cite{li2003improving}. To partition high dimensional and large data space, a grid-based algorithm was proposed in\,\cite{wei2007grid}. The algorithm works in two phases. Firstly, it creates the non-overlapping d-dimensional cells using the domain space followed by  partition-based clustering. The proposed approach led to a high detection rate and a relatively low false-positive rate.

\begin{table*}[tbh]

\caption{Summary of Intrusion Detection in CPS using Unsupervised Learning techniques}
\label{tab:UL1}
\resizebox{18.15cm}{10.2cm}{
\begin{tabular}{C{1cm}C{1.5cm}C{1.5cm}C{1.75cm}C{2cm}C{1.5cm}C{1cm}C{1.5cm}C{1.4cm}C{1.4cm}C{1.7cm}}
\hline
\textbf{Work} & \textbf{Domain} & \textbf{Audit Material} & \textbf{Complexity} & \textbf{Algorithms} & \textbf{Feature Selection} & \textbf{Time Series} & \textbf{Dataset} & \textbf{Data Type} & \textbf{Data Available} & \textbf{Metrics} \\ \hline \hline
\cite{liu2014practical} & Annon & Network & Simple & k-means, and SVM & No & Yes & Annon & Actual & No & ADR, and Accuracy \\ 
\cite{pal2017effectiveness} & Water & Physical & Simple & Apriori & No & Yes & SWaT & Actual & Yes & Accuracy \\ 
\cite{nader2014norms,nader2014mahalanobis} & Gas, and Water & Physical & Simple & OCSVM & No & No & MSU & Actual & Yes & Accuracy \\ 
\cite{krishnamurthy2014scalable} & HVAC & Hybrid & Simple & BN & No & No & Annon & Actual & No & Accuracy \\ 
\cite{mantere2014module, mantere2013network} & Printed Intelligence & Network & Hybrid & SOM & No & No & PrintoCent & Actual & No & None \\ 
\cite{ahmed2019unsupervised} & Energy & Physical & Simple & IF, PCA, SVM,k-NN, NB, and MLP & No & Annon & SE-MF & Actual & No & Accuracy, and F1 score \\ 
\cite{alves2018embedding} & Water & Network & Simple & k-means, and LOF & Yes & Yes & Annon & Actual & No & PLC Scan Time \\ 
\cite{pasricha2017special} & Annon & Network, and Physical & Hybrid & k-NN, SVM, SVR, and AR & Yes & Yes & Annon & Actual & No & SR, and NFAR \\ 
\cite{linda2009neural} & Water & Network & Simple & NN & Yes & Yes & Annon & Actual & No & Recall, and FPR \\ 
\cite{khalili2015sysdetect} & Water & Physical & Simple & Apriori & No & No & Annon & Actual & No & Accuracy \\ 
\cite{hadvziosmanovic2012n} & Annon & Network & Simple & PAYL, POSEIDON, Anagram, McPAD & No & No & Annon & Actual & No & FPR, and DR \\ 
\cite{shin2010experimental} & Annon & Network & Simple & Multi Hop Clustering & No & No & Annon & Simulated & No & DR \\ 
\cite{lauf2010distributed} & Aviation, and Robots & Network & Hybrid & Statistical & No & No & Annon & Simulated & No & \% of Devient Nodes for convergence \\ 
\cite{kwon2015behavior} & Energy & Network & Simple & Statistical & No & No & Korean substation & Actual & No & Precision, Recall, F1 score, FPR, and FNR \\ 
\cite{sayegh2014scada} & Energy & Network & Hybrid & Bayesian & No & Yes & American University of Beirut power plant & Actual & No & Accuracy, and FP \\ 
\cite{nader2013intrusion} & Gas & Physical & Simple & OCSVM & No & No & MSU & Actual & Yes & Accuracy \\ 
\cite{umer2017integrating, umer2020generating,ahmed2021machine} & Water & Physical & Simple & FP-growth & Yes & Yes & SWaT & Actual & Yes & Accuracy \\ 
\cite{bhattacharjee2018towards} & Energy & Physical & Simple & k-means & No & Yes & PeCanStreet Project, and Irish Social Science Data Archive & Actual & Yes & Accuracy \\ 
\cite{dussel2009cyber} & SCADA & Network & Simple & Statistical & No & No & AUT09 & Actual & No & DR, and FP \\ 
\cite{yang2006anomaly} & SCADA & Hybrid & Simple & Statistical & Yes & Yes & Annon & Simulated & No & Detection Diagrams \\ 
\cite{maglaras2014intrusion} & SCADA & Network & Simple & OCSVM & No & No & Annon & Actual & No & Accuracy \\ \hline
\end{tabular}
}

\end{table*}

\subsubsection{Association Rule Mining}
ARM\,\cite{agrawal1993mining} is a rule-based machine learning technique used to uncover relationships in databases. Traditionally, it was used for market basket analysis. It has several applications such as predicting customer behavior, product clustering, web usage mining, catalog design, store layout, bioinformatics, and intrusion detection. 

ARM works on the principle of Support and Confidence. Support is calculated using the itemset. An itemset is a set of values of one or more attributes. Itemsets that meet the support threshold are called as frequent itemsets. Support for an item set $A$ in $D$ can be defined as the proportion of examples (rows, or transactions) $e$ in the dataset that contains $A$. Formally, it can be defined as follows:
\begin{equation}
\textstyle{S}(A) = \frac{|e \in D; A \in e|} {|D|}
\end{equation}

Confidence is the proportion of rules that contain both the antecedent and  the consequent. It measures the frequency of the rule w.r.t. the antecedent. The confidence of $X \implies Y$ can be defined as follows:

\begin{equation}
\textstyle{C}(X \;\implies\; Y) = \frac{S(X \cup Y)} {S(X)}.
\end{equation}

Frequent itemsets are partitioned in one or more ways to generate rule such as  $X \implies Y$, where $X$ is antecedent, and $Y$ the consequent. Rules that satisfy the confidence threshold are qualified for the final set of association rules.

ARM was used in\,\cite{khalili2015sysdetect} to determine the critical system state for the intrusion detection system using the Apriori algorithm.  At the same time, it also incorporated the expert opinion for the identification of critical states. The expert opinion was used in each iteration to reduce the number of candidates in the following iteration. ARM was also used in\,\cite{pal2017effectiveness} to generate  invariants  for a water treatment plant using the Apriori algorithm. This was a preliminary work to discuss the effectiveness of ARM as a proof of concept. It only mined the rules, or invariants, for pairwise sensors/actuators. Secondly, the accuracy of the proposed approach was not effective for practical implementation due to False Positives and False Negatives. Subsequently in \,\cite{umer2017integrating, umer2020generating, ahmed2021machine} invariants were mined on the same plant using the FP-Growth algorithm. The approach succeeded in mining a more exhaustive set of invariants including local and global invariants. Here, ``local" refers to within a process and ``global" to inter-process invariants. The invariants mined are available at\,\cite{swatDataset}. The invariants were also placed a monitors for distributed attack detection in the plant. The accuracy of the proposed approach was  promising considering that the implementation was on an operational plant.

\begin{figure*}[tbh]
\centering
\includegraphics[width=1.0\linewidth,height=7.5cm]{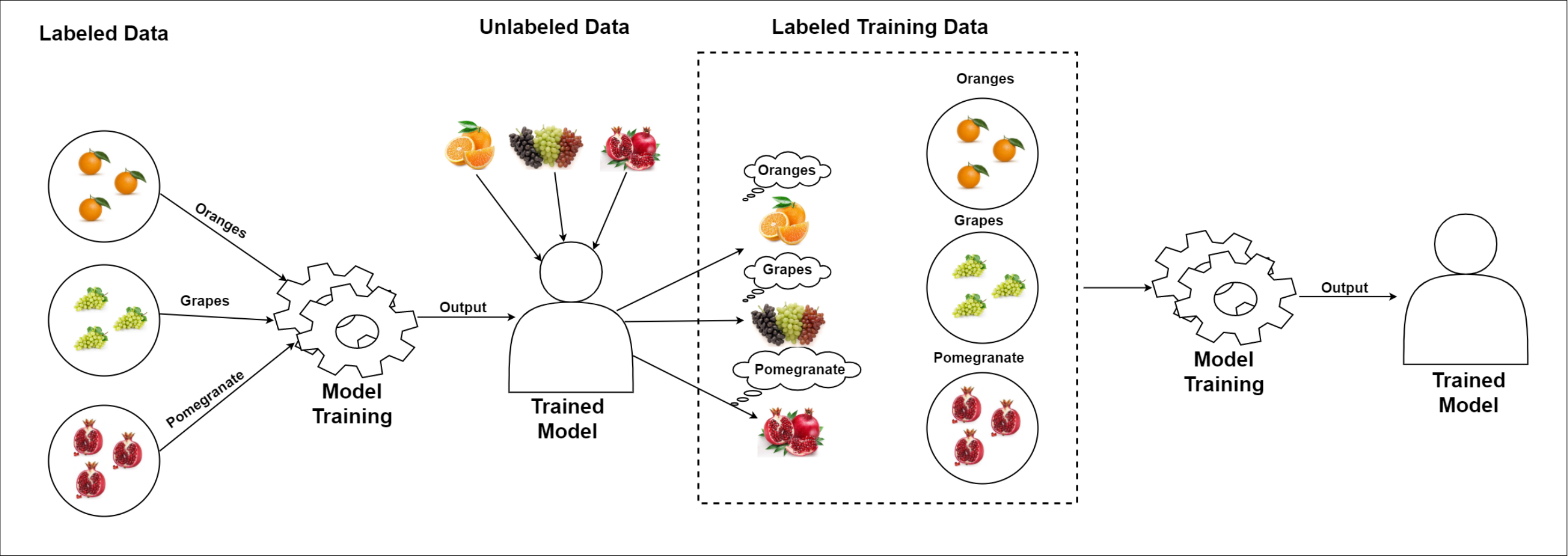}
\caption{Semi-Supervised Learning}
\label{fig:SSL}
\end{figure*}

\subsection{Deep Learning Based Unsupervised Learning Approaches}
\label{DLUL}

There exist several unsupervised deep learning approaches though only a few studies have been reported for securing ICS. Events originating between the application layer to the kernel layer get recorded in system logs and traces. These logs and traces are  helpful in monitoring the performance of any system and  are useful for anomaly detection. However, these traces and logs are generally large in a real-time system,  and therefore online anomaly detection remains a challenge for such systems. A deep recursive attentive model (DReAM) was proposed in\,\cite{ezeme2019dream} to detect anomalies through temporal information of the system using execution sequences. DReAM works on two components, namely,   the unsupervised recurrent neural network predictor and the supervised clustering classifier. Similarly, Mobile Edge Computing (MEC) aims to do intensive computation at the edge networks. This has led to an increase in traffic of transportation networks and the key security issues as well. Therefore, a DL-based framework was proposed in\,\cite{chen2019deep} using DBN to learn the model. Its performance was compared with traditional ML-based algorithms. The proposed method was able to detect attacks with acceptable accuracy, but with higher time complexity thus rendering it unsuitable for streaming data. DL-based clustering techniques are further discussed in the following subsections.

\subsubsection{Autoencoder based deep clustering}
Autoencoder (AE) is a type of ANN that works in an unsupervised manner to learn efficient data encodings\,\cite{kramer1991nonlinear}. It first learns the representation, i.e., encoding from data and then used for dimensionality reduction. It thus trains the network to ignore  noise. It tries to learn a representation close enough to the original input while minimizing the reconstruction loss. There are several AE-based deep clustering methods including Deep Clustering Network (DCN)\,\cite{ yang2017towards}, Deep Embedding Network (DEN)\,\cite{ huang2014deep}, Deep Subspace Clustering Networks (DSC-Nets)\,\cite{ peng2017deep}, Deep Multi-Manifold Clustering (DMC)\,\cite{ chen2017unsupervised}, Deep Embedded Regularized Clustering (DEPICT)\,\cite{ghasedi2017deep}, and Deep Continuous Clustering (DCC)\,\cite{ shah2018deep}.

\subsubsection{Clustering Deep Neural Network (CDNN)}
This method trains the model primarily on clustering loss. Therefore, if reconstruction loss is not properly designed, then it may lead to a corrupted feature space. Based on network initialization, it can be classified into unsupervised pre-trained, supervised pre-trained, and non-pre-trained network\,\cite{min2018survey}.

\subsubsection{Variational Autoencoder (VAE)-based deep clustering}
In VAE, the latent code of AE is bound to follow a predefined distribution. It is a combination of Bayesian methods\,\cite{min2018survey}. It can use stochastic gradient descent\,\cite{bottou2010large} and standard backpropagation\,\cite{hecht1992theory} to optimize the variational inference.

\subsubsection{Generative Adversarial Network (GAN)-based deep clustering}
GAN-based clustering works on the principle of the min-max adversarial game.  Two types of networks are used, namely,  generative  and  discriminative\,\cite{min2018survey}. The generative network attempts to map a sample from prior distribution to data space whereas the discriminative network maps the input as a real sample of the distribution by computing the probability. There are various GAN-based deep clustering algorithms including Deep Adversarial Clustering (DAC)\,\cite{harchaoui2017deep}, Categorical Generative Adversarial Network
(CatGAN)\,\cite{springenberg2015unsupervised}, and Information Maximizing Generative Adversarial
Network (InfoGAN)\,\cite{chen2016infogan}.

\begin{figure*}[tbh]
\centering
\includegraphics[width=0.5\linewidth, height=5cm]{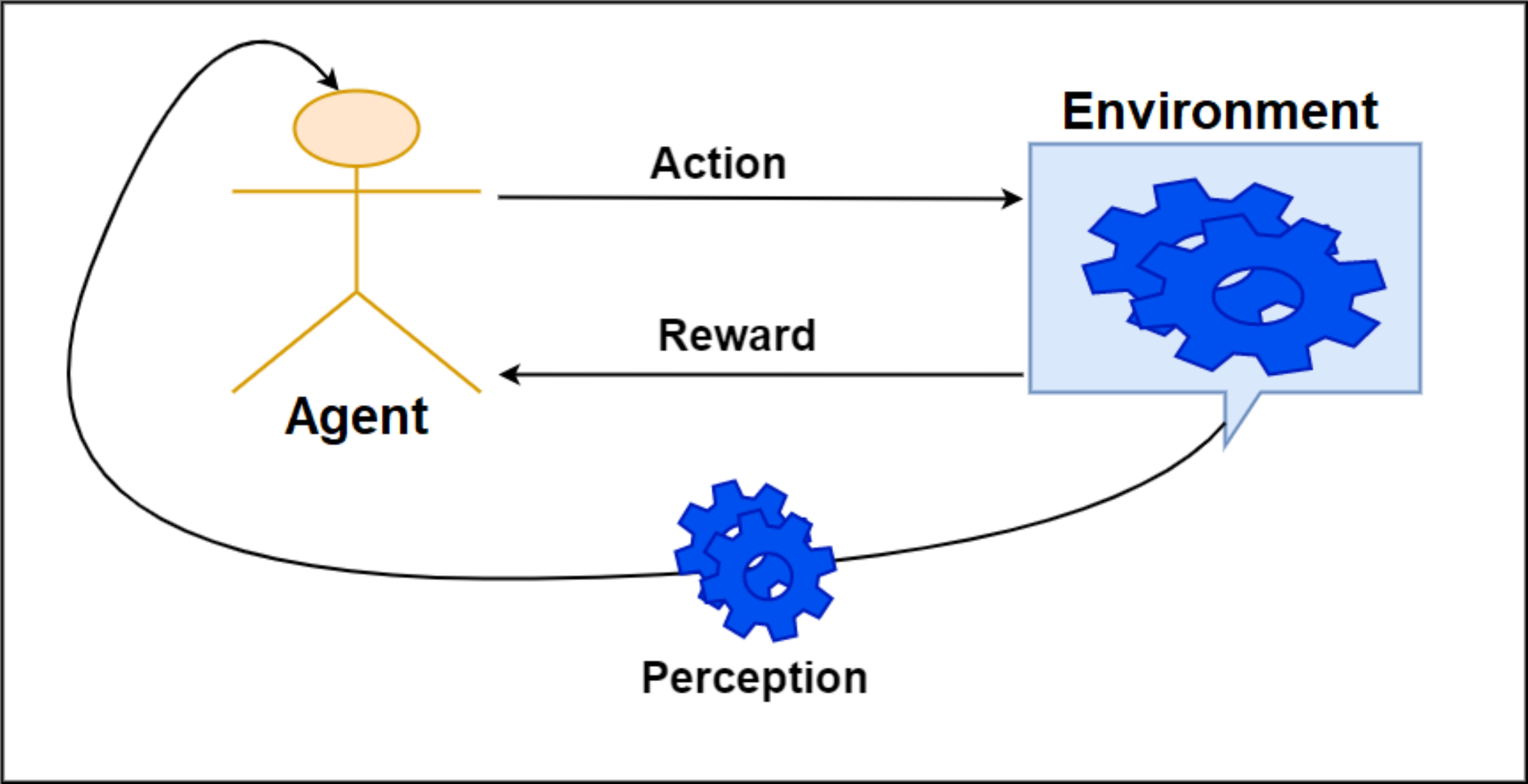}
\caption{Reinforcement Learning}
\label{fig:RL}
\end{figure*}
\section{Semi-Supervised Learning}
\label{sec:semisupervisedLearning}
Semi-Supervised Learning (SSL) uses both the labeled and unlabeled data for training of the model, one way of doing SSL is described in Figure\,\ref{fig:SSL}. In the first phase,  the model is trained using the labeled data as in supervised learning. In the second phase, it assigns the labels to the unlabeled data using the model trained in the earlier phase. In the third phase,  both the initially given labeled data and the newly assigned labeled data  are used for  training  the model. The following assumptions are made to label the unlabeled data.

\subsection{Continuity assumption}
This assumption works on the principle  that points closer to each other are likely to share the same label. This assumption is also used in SL to create decision boundaries. In SSL, this assumption prefers decision boundaries that are in lower density regions. Thus, it is possible that some points are close to each other but may lie in different classes.

\subsection{Cluster assumption}
This assumption considers that data points are scattered across  clusters. The data points present in the same cluster should share the same label.

\subsection{Manifold assumption}
In this assumption data points lie on a manifold of a lower dimension as compared to the input space. This assumption can eliminate the curse of dimensionality if the manifold is learned using both labeled and unlabeled data. Further learning can be done using distances and densities set out on manifold. 

SSL approaches are worth exploring. They have exhibited performance better than supervised and unsupervised approaches when the size of labeled data is relatively small\,\cite{criminisi2012decision,junejo2013robust,luo2013manifold}. A study reported in\,\cite{huda2019automatic} proposed a model to extract the behavioral patterns of malware using semi-supervised and unsupervised machine learning techniques. SSL was used in\,\cite{huda2017defending} to automatically update the attack detection system of CPS using the unlabeled malware data. At the first stage, it captures malware patterns from unlabeled data using UL. Then, this information is used by the classification system of the detection engine. The proposed approach used the k-means for clustering and SVM for classification. SSL is also used for fault detection as in\,\cite{zhang2017fault} using Local Linear Embedding (LLE). LLE is usually used for fault detection in ICS. It only preserves the local information of the structure while ignores the global properties of data. The proposed approach integrated SSL into LLE to utilize the labeled data. The studies reported in\,\cite{symons2012nonparametric,wagh2014effective,gao2013improved} have shown that semi-supervised approaches can perform better than supervised and unsupervised approaches for conventional network intrusion detection, but these studies are yet to make their way through to IDS for ICS.

\section{Reinforcement Learning}
\label{RL}\label{sec:reinforcementLearning}
Reinforcement Learning (RL) is  significantly different from  other ML techniques. In RL there exist three  main components of the learning system, namely, an Agent, the Environment, and Reward.  As illustrated in Figure\,\ref{fig:RL}, an agent performs an action in the environment for which it receives  reward, which could be positive or negative. This way the agent learns in the environment. RL does not require a dataset for learning as required in other ML techniques. Some commonly used RL algorithms are introduced next.

\subsection{Temporal difference (TD) learning}
TD is a model-free learning algorithm. The model is learned using bootstrapping done  using the current estimate of the value function. Methods are sampled from the environment and updates are performed based on the current estimate\,\cite{sutton1998introduction}.

\subsection{State–action–reward–state–action (SARSA)}
SARSA learns using a Markov Decision Process. It performs actions on the environment and updates its policy based on the reward received against those actions. Initial conditions, learning rate, and discount factor are the hyper-parameters of the algorithm.

\subsection{Q-learning}
Q-learning is also a model-free algorithm and does not require a model of the environment. It lets the agent learn a policy to perform an action based on different circumstances. It does not require adaptations because of stochastic transitions and rewards.

RL was used in\,\cite{otoum2019empowering} for intrusion detection in a simulated Wireless Sensor Network (WSN) environment. The authors also  compared their  work with adaptive ML-based IDS. RL-based IDS performed better than other ML-based IDS. A model-free based RL approach was proposed in\,\cite{kurt2018online} for anomaly detection in the smart grid. They proposed an RL based solution to the Partially Observable Markov Decision Process (POMDP) problem. For the optimal defense of CPS in\,\cite{feng2017deep}, the problem was formulated as a two-player zero-sum game. Deep RL was used to tune the actor-critic Neural Network structure. Likewise in\,\cite{panfili2018game},  a multi-agent general sum game  was used to model the attack problem.  RL  was used to find the optimal solution for prevention actions and the associated costs. A proof-of-concept was  provided  by simulating a subsystem of the ATENA controller\,\cite{atena}. Q-Learning based vulnerability assessment of smart grid is reported in\,\cite{yan2016q} where sequential topological attacks were the targets. Using Q-Learning, an attacker can cause severe damage  to a plant. The effectiveness of the proposed approach was demonstrated using IEEE 5-bus, RTS-79, and IEEE 300-bus systems-based simulation results. RL was also used in\,\cite{lu2017motor} for anomaly detection in Unmanned Aerial Vehicle (UAV). It recorded the temperature of the motor using sensors and  used a Raspberry Pi based processing unit to observe the anomalous behavior of the motor.

\section{Major Challenges and Recommendations for IDS in ICS}
\label{sec:challenges}
\subsection{Adversarial Machine Learning for IDS}
Machine learning is being used for intrusion detection, at the same time Adversarial machine learning is being used to counter its benefits. For example, in\,\cite{zizzo2019adversarial} LSTM was used to train the model on the normal data from a real-world ICS and its performance tested on attack data. Further, the adaptive attacks were performed to deteriorate the performance of the machine learning classifier. Machine Learning as a service (MLaas) is also gaining popularity in cloud-based services. They typically use deep neural networks (DNN) for different predictive models. Now they have become vulnerable to different adversarial attacks. In this case, the adversary try to steal the model by querying the Application Programming Interface (API). For example a study proposed in \cite{yu2020cloudleak} used an attack methodology to extract the DNN models from various cloud-based platforms. For this purpose, they used various algorithms including active and transfer learning. Similarly, \cite{lin2020composite} used composite attacks using Trojan triggers to disrupt the performance of DNN model. Their Trojan triggers were composed of benign features of multiple labels. The model misclassifies the output when input is stamped with Trojan trigger. 

There are some studies which have tried to tackle this challenge but still more work needs to be done. For example, \cite{zhang2020zero} used the zero knowledge proofs for decision tree. They reported their accuracy and predictions on public dataset without leaking any information about the model. Using the proposed study a decision tree having depth of 23 levels and 1029 number of nodes can generate the zero knowledge proofs in 250 seconds. Likewise, \cite{li2020deepdyve} used simple and smaller pre-trained neural network models for the verification of DNN-based systems and to protect them against adversarial attacks.  We believe that these types of approaches could be useful for defending the adversarial attacks on machine learning models.

\subsection{Lack of Attack Patterns and its Mitigation in ICS}
It is difficult to produce an exhaustive dictionary of attack signature in complex physical processes in ICS. Therefore it becomes difficult to detect zero-day attacks. For example, the model proposed in\,\cite{nahmias2019trustsign} is robust and fast as it does not require training on new input data. However, as it generates signatures using only the available malware processes, it could be prone to zero-day attacks. The study reported in \cite{umer2021attack} used the unsupervised machine learning approach to generate attacks for a real-world ICS. They used association rule mining to generate attack patterns. Normally, Supervised learning approaches lacks attack data, therefore the study reported in \cite{umer2021attack} could be beneficial for making robust supervised learning-based IDS. Moreover, it can also be useful for signature-based approaches for IDS, as it automatically generates the signatures using the attacked data on real-world ICS. Cyber-attacks were modeled as timed-automaton in \cite{SUGUMAR2019100324} for SWaT \cite{7469060}. This model was used as a baseline to create a number of cyber-attacks using mutation. Though all the created attacks may not be actual attacks but it seems to be a good strategy to defend against zero-day attacks because of the comprehensive attack dictionary created by the proposed approach. Similarly, a study reported in \cite{JIA2021100452} used a gradient-based attack scheme to generate attacks for real-world ICS. Through their approach they mislead the RNN based anomaly detector of two real-world ICS namely SWaT \cite{7469060} and WADI \cite{mujeebPalletiMathur}. 

\subsection{Aging and Complexity of the Physical Systems in ICS}
There are serious issues related to the aging and complexity of the physical systems while dealing with specification-based approach. There could be inaccuracies in the operational manuals, and interpretation of the process behavior. Though behavior-based approach is favoured against incorrect vendor specifications due to its dependability on empirical data but there are issues of detecting zero-day attacks, ensuring an acceptable rate of false alarms, and managing computational complexity


\subsection{Heterogeneity among Physical Processes in ICS}
There exists a heterogeneous behavior among physical processes of an ICS because components, size, and functionality of each process is different from others \cite{7469060,swatDataset}. Therefore it is a challenge to detect attacks in the heterogeneous physical processes controlled by an ICS. For example, SWaT testbed \cite{swatDataset} which is an industrial scaled-down replica of a water treatment plant has different six stages. Here attack on one stage can disrupt the processes of other stages as well. So developing a model which can capture the behavior of heterogeneous physical processes is still a major challenge. Though IDS based on physical process have received relatively little attention but now there is a growing trend to detect intrusion at the physical process level.

\subsection{Inherent Class Imbalance Nature of IDS}
Behavior-based approaches suffer from skewed class problems. Here most of the data belongs only to a single class (normal behavior). Any naive classifier that labels each instance as normal will get a higher accuracy. Therefore, accuracy is not enough to assess the performance of IDS. It is also important to note that acceptable values of the metrics discussed in section \ref{metrics} might still not make an IDS suitable for deployment in an operational plant. As an example, consider accuracy. A high accuracy can be obtained by having high values of TP and TN and relatively lower values of FP and FN. However, suppose that accuracy is high, lets say, 99\%, but the number of false positives (FP) per day is, say, 50. Such an IDS would likely be not used in an operational plant. Thus, it is recommended that in addition to reporting one or more metrics mentioned above, FP must also be reported to assess how well an IDS would perform when deployed in a constantly running plant.

\subsection{Stealthy Attacks on ICS}
If the attacker has deep insights of the system then it would be vulnerable to stealthy attacks. These types of attacks  gradually disrupt the performance of the operational plant. Though there are a number of studies on this issue but the detection of these attacks is still a major challenge. For example, a study proposed in \cite{sun2020detecting} used the Profile-DNS for detecting the stealthy attacks by characterizing the expected DNS behavior. Likewise, \cite{9312437} used VMshield for securing the cloud platforms against the stealthy attacks. They did feature selection using meta-heuristic , and binary particle swarm optimization (BPSO) algorithms. They used Random Forest for the classification of malicious and benign processes. 

\subsection{Association Rule Mining for IDS}
Most of the reported ML-based intrusion detection work in ICS   uses SL approaches while there exists only a sprinkling  of work  using UL approaches. Particularly, only a few studies have reported the use of an ARM-based UL approach for intrusion detection in ICS\,\cite{umer2017integrating, umer2020generating,pal2017effectiveness,khalili2015sysdetect}. Despite this,  there remain  gaps that need to be filled. For example, all the studies reported in \cite{umer2017integrating, umer2020generating,pal2017effectiveness,khalili2015sysdetect} used data from  an ICS controlling a water plant. Though,\,\cite{umer2020generating} practically implemented the ARM-approach in an operational plant with  promising results, the same approach  needs to be tested on other ICS  used in systems such as the smart grid and  gas plants. Moreover, \cite{umer2020generating} used a time series data but used the FP-Growth algorithm to mine the rules. FP-Growth is time agnostic, therefore could be promising to use Temporal Association Rule Mining\,\cite{liang2005temporal} for that purpose.
\subsection{Deep Learning for IDS}
Deep learning can be an effective approach for detecting anomalies in ICS-controlled plants.   It can automatically generate features based on the physical dynamics of each ICS. Some studies reported the use of DL to secure ICS as discussed in section\,\ref{DLSL} and\,\ref{DLUL}  most of which are SL approaches. There exist only a few studies\,\cite{chen2019deep, ezeme2019dream} where the UL approach is used. There are several DL-based clustering techniques as discussed in section\,\ref{DLUL} that need to be explored for securing the ICS. However, higher time complexity possesses a great challenge for their application in real-time systems, such as ICS.

\subsection{Agent-based Learning for Securing ICS}
Reinforcement Learning which works on the basis of agent and environment interaction is the least explored area for securing ICS. Though there are a few studies reported in the literature as discussed in section\,\ref{RL}. However, considering the dynamic nature of ICS in different domains including smart grids, water, gas, etc,  RL appears a  promising avenue to explore and implement  in various domains.

\subsection{Zero-shot Learning for Resilience against Zero-day Attacks}
Apart from the ones mentioned above, there are  other promising ML approaches that need to be explored for intrusion detection in ICS. Zero-shot learning \cite{romera2015embarrassingly,socher2013zero} is one such promising approach for detecting zero-day attacks. Domain adaptation can help learn an IDS for one ICS using data of some other ICS. Lastly, distribution shift techniques can be explored for making the model adapt to the changing behavior of the system with time. The above-mentioned approaches remain to be explored in depth to effectively solve the problem of intrusion detection.

\subsection{Comparative Analysis of Behavior-based Approach with Specification and Signature-based Approach for IDS}
Even though some studies compare various ML algorithms on their dataset, we were not able to find a comparison with the specification or signature-based techniques except in\,\cite{umer2017integrating, umer2020generating} where  the behavior-based approach  was compared with the specification based approach. A comparison of the three types of approaches needs to be conducted on the same dataset under similar assumptions to gain a better understanding of their effectiveness in detecting cyber-attacks.

\subsection{Need of Comprehensive Evaluation Metrics for Real-world ICS}
Several studies have reported only a few metrics such as either accuracy (or error rate/graphs) or detection rates as discussed in section \ref{metrics}. Reporting only one or two performance metrics for a skewed class problem is not sufficient, more than one of the following metrics should be used: accuracy, precision, recall, F-measure, ROC, and AUC. Only a few studies have reported AUC or ROC despite the fact that these are more appropriate measure of classifier performance in IDS. Secondly, there is little focus in the literature to report the time to detect an attack or the percentage detection of an attack over the duration for which it lasts. The use of these measures should be made more prevalent.

\subsection{Multi-layered Defense for IDS}
A majority of the approaches  focus on detecting intrusion at the network layer. After all, this is the first line of defense of an ICS, though often  easier to breach as many ICS are using ready-made industrial protocols, and due to insider threats. Once breached, detecting intrusions in the physical layer improves the chances of avoiding plant damage or service disruption.  Detecting cyber-attacks at this layer would be more promising as each ICS is unique and to be successful,  the attacker would require a knowledge of the physical dynamics of that particular ICS. Therefore more attention seems necessary in developing IDS for the physical layer consisting of at least a few dozen sensors and actuator attributes. We believe that the final solution lies in a multi-layered defense, a network IDS followed by a physical process IDS.

\subsection{Root Cause Isolation for IDS}
While detecting a cyber-attack launched by an intruder  is the primary goal of an IDS,  detecting the nature and location of the ongoing attack, and taking further actions,   remain  crucial to steps. Only one work\,\cite{krishnamurthy2014scalable} has reported  root cause isolation. Lastly, few studies have modeled the problem as a time series problem, whereas, many ICS repeat the same operations over and over again. More approaches are needed to address these issues.

\section{Conclusion}
\label{sec:conclusions}
ICS are critical for the economy and infrastructure of any country and hence ought to be protected against cyber-adversaries. These adversaries could be  hackers, enemy states, and displeased employees, etc. Therefore securing an ICS from cyber-attacks is one of the prime concerns for governments and organizations. Behavior-based approaches such as machine learning, deep learning, and statistical approaches for intrusion detection, are gaining attention. They can be automated, several scale well, and can be generalized  and are becoming affordable to apply because of cheaper and widely available computational power. Thus, this survey focuses on  literature to consolidate the work on behavior-based approaches for IDS in ICS, categorizes them, identifies gaps, and proposes future research directions. This area This area is in a  need of a high fidelity benchmark dataset. There is  room to apply the newly developed ML techniques and compare them with the specification and  signature-based approaches, especially for the physical process controlled by  an ICS. Time series modeling of the problem and the use of new metrics is also required. All in all, ML and DL approaches are promising techniques for the detection of cyber-attacks  in both the network and physical process layer of an ICS, though there is  room for improvement.

\section*{Acknowledgements}
The authors acknowledge the time and efforts made by Mr. Bilal Hayat Butt for providing valuable suggestions and feedback on the survey.

\def\abbrevOne{\hbox{
\begin{tabular}{|l|p{0.38\textwidth}|}
\hline
\multicolumn{1}{|c|}{\textbf{Term$^*$}} &\multicolumn{1}{c|}{\textbf{Expansion}}   \\ \hline
\mlterm{AE}&Autoencoder\\
AMI&Advanced Metering Infrastructure\\
\mlterm{ANN}& Artificial Neural Networks\\
\mlterm{ARM}&Association Rule Mining\\
\mlterm{AUC}&Area Under ROC\\
BACnet& Building Automation Control Network\\
\mlterm{BayesLR}&Bayes Logistic Regression\\
\mlterm{BayesNet}&Bayes Network\\
\mlterm{BFTree}&Best First Tree\\
\mlterm{CatGAN}&Categorical Generative Adversarial Network \\
\mlterm{CDNN}&Clustering Deep Neural Network\\
\mlterm{CNN}&Convolutional Neural Networks\\
CPS&Cyber-Physical System\\
\mlterm{DAC}&Deep Adversarial Clustering\\
\mlterm{DBN}&Deep Belief Networks Decision Process \\
\mlterm{DCC}&Deep Continuous Clustering\\
\mlterm{DEN}&Deep Embedding Network\\
\mlterm{DEPICT}&Deep Embedded Regularized Clustering\\
\mlterm{DL}&Deep learning\\
\mlterm{DMC}&Deep Multi-Manifold Clustering\\
DNP3&Distributed Network Protocol 3 \\
\mlterm{DReAM}&Deep Recursive Attentive Model\\
\mlterm{DSC-Nets}&Deep Subspace Clustering Networks\\
\mlterm{FN}&False Negative\\
\mlterm{FP}& False Positive\\
\mlterm{GAN}&Generative Adversarial Network\\
ICMP&Internet Control Message Protocol\\
ICS& Industrial Control Systems \\
IDS&Intrusion Detection System\\
\mlterm{InfoGAN}&Information Maximizing Generative Adversarial Network \\
\mlterm{ML}& machine learning\\
LAN&Local Area Network\\
\hline
\end{tabular}
}}

\def\abbrevTwo{\hbox{
\begin{tabular}{|l|p{0.38\textwidth}|}
\hline
\multicolumn{1}{|c|}{\textbf{Term$^*$}} &\multicolumn{1}{c|}{\textbf{Expansion}}  \\ \hline

\mlterm{LDA}&Linear Discriminent Analysis\\
\mlterm{LLE}&Local Linear Embedding\\
\mlterm{LR}&Logistic Regression\\
\mlterm{LSTM}&Long-Short Term Memory\\
\mlterm{MLP}&Multi-Layer Perceptron\\
\mlterm{NB}&Naive Bayes\\
\mlterm{NNGE}&Non-Nested Generalized Exemplers \\
\mlterm{OCC}&One-Class Classification\\
\mlterm{OneR}&One Rule\\
PLC&Programmable Logic Controller\\
PMU&Phasor  Management Unit\\
\mlterm{POMDP}&Partially Observable Markov Decision Process \\
\mlterm{RL}&Reinforcement Learning\\
\mlterm{RF}&Random Forest\\
\mlterm{RNN}&Recurrent Neural Networks\\
\mlterm{ROC}&Receiver Operating Characteristic\\
\mlterm{SAE}&Stacked Autoencoder\\
\mlterm{SARSA}&State–action–reward–state–action \\
SCADA&Supervisory Control and data Acquisition System\\
\mlterm{SL}&Supervised Learning\\
\mlterm{SOM}&Self-Organizing Maps\\
\mlterm{SVM}&Support Vector Machine\\
\mlterm{SSL}&Semi-Supervised Learning\\
TCP& Transmission Control Protocol\\
\mlterm{TD}&Temporal difference \\
\mlterm{TP, TPR}&True Positive, True Positive Rate\\
\mlterm{TN}&True Negative\\
UAV&Unmanned Aerial Vehicle\\UDP&User Datagram Protocol\\
\mlterm{UL}&Unsupervised Learning\\
\mlterm{VAE}&Variational Autoencoder\\
&\\
\hline
\end{tabular}
}}

\begin{table*}[tbh]
\caption{Abbreviations used in the survey.$^*$Terms in bold  are used  in machine learning literature.}
\label{tab:abbreviations}
\def\tableHeight{0.68}
\begin{tabular}{l }
\vbox to \tableHeight\textheight{\abbrevOne\vfill}\vbox to \tableHeight\textheight{\abbrevTwo\vfill}\\
\end{tabular}
\end{table*}


\bibliographystyle{ACM-Reference-Format}
\bibliography{references}


\end{document}